%% file: main.tex
\documentclass[
reprint,
superscriptaddress,
amsmath,
amssymb,
aps,
prl,
floatfix
]{revtex4-2}

\usepackage{graphicx}
\usepackage{dcolumn}
\usepackage{bm}
\usepackage[colorlinks,citecolor=magenta]{hyperref}

\begin{document}

\title{Probing Dark Excitons in Monolayer MoS$_2$ by NonLinear Two-Photon Spectroscopy}

\author{Chenjiang Qian}
\email{chenjiang.qian@iphy.ac.cn}
\affiliation{Walter Schottky Institut and TUM School of Natural Sciences, Technische Universit{\" a}t M{\" u}nchen, Am Coulombwall 4, 85748 Garching, Germany}
\affiliation{Beijing National Laboratory for Condensed Matter Physics, Institute of Physics, Chinese Academy of Sciences, Beijing 100190, China}
\affiliation{School of Physical Sciences, University of Chinese Academy of Sciences, Beijing 100049, China}
\author{Viviana Villafañe}
\affiliation{Walter Schottky Institut and TUM School of Natural Sciences, Technische Universit{\" a}t M{\" u}nchen, Am Coulombwall 4, 85748 Garching, Germany}
\author{Pedro Soubelet}
\affiliation{Walter Schottky Institut and TUM School of Natural Sciences, Technische Universit{\" a}t M{\" u}nchen, Am Coulombwall 4, 85748 Garching, Germany}
\author{Peirui Ji}
\affiliation{Walter Schottky Institut and TUM School of Natural Sciences, Technische Universit{\" a}t M{\" u}nchen, Am Coulombwall 4, 85748 Garching, Germany}
\author{Andreas V. Stier}
\affiliation{Walter Schottky Institut and TUM School of Natural Sciences, Technische Universit{\" a}t M{\" u}nchen, Am Coulombwall 4, 85748 Garching, Germany}
\author{Jonathan J. Finley}
\email{finley@wsi.tum.de}
\affiliation{Walter Schottky Institut and TUM School of Natural Sciences, Technische Universit{\" a}t M{\" u}nchen, Am Coulombwall 4, 85748 Garching, Germany}

\begin{abstract}

We report a new dark exciton in monolayer MoS$_2$ using second harmonic generation spectroscopy.
Hereby, the spectrally dependent second harmonic generation intensity splits into two branches, and an anticrossing is observed at $\sim$ 25 meV blue detuned from the bright neutral exciton.
These observations are indicative of coherent quantum interference arising from strong two-photon light-matter interaction with an excitonic state that is dark for single photon interaction.
The existence of the dark state is supported by engineering its relaxation to bright localized excitons, mediated by vibrational modes of a proximal nanobeam cavity.
We show that two-photon light-matter interaction involving dark states has the potential to control relaxation pathways induced by nanostructuring the local environment.
Moreover, our results indicate that dark excitons have significant potential for nonlinear quantum devices based on their nontrivial excitonic photophysics.
\end{abstract}

\maketitle


Nonlinear phenomena play a central role in advanced photonic and electronic devices since they govern functionalities such as transduction and switching and lead to nontrivial physics in which different degrees of freedom interact \cite{stegeman2012}.
Two-dimensional (2D) semiconductors are an ideal platform to investigate nonlinear optical phenomena since they not only have strong $\chi^{2}$ coefficient for second harmonic generation (SHG) \cite{10.1002/adma.201705963,doi.org/10.1002/inf2.12024,10.1038/nphoton.2012.147} but also possess rich exciton photophysics with electronic states that couple to both photons and phonons in the system \cite{10.1038/nnano.2015.73,PhysRevLett.114.097403,10.1038/s41467-021-27213-8}.
As such, nonlinear spectroscopy of 2D semiconductors and their heterostructures currently attracts broad attention in the context of advanced nonlinear optoelectronic devices, as well as being an important tool to probe fundamental electronic excitations in the material \cite{10.1038/s41566-021-00859-y,PhysRevB.105.115420,10.1002/lpor.202100726}.

The multiphoton feature in nonlinear optics provides parity control in the light-matter interaction since the coupling operator $\propto\hat{r}^N$, where $\hat{r}$ represents the position operator, has opposite symmetry for odd or even photon numbers $N$ \cite{PhysRevLett.130.083602,PhysRevLett.122.043601,PhysRevLett.72.1001}.
This means that dipole forbidden (dark) excitons having odd parity for single-photon coupling ($N=1$) can become bright for the two-photon ($N=2$) coupling \cite{10.1038/nature13734,PhysRevB.88.085321}, and vice versa.
Usually, two-photon strong coupling for bright excitons in semiconductors is obtained for systems where two real (excitonic) states are mutually close to resonance with one and two photon transitions.
For example, this occurs for self-assembled quantum dots having the exciton and biexciton states, or for 2D semiconductors having multiple electronic levels resonant to different number of photons \cite{10.1038/s41567-018-0384-5,PhysRevLett.120.213901,10.1038/s41467-021-21547-z,10.1364/CLEO_FS.2023.FW4N.4}.
In contrast, dark excitons have the potential for direct, coherent two-photon light-matter interaction without an intermediate one-photon resonance.
This opens the way for the coherent control of dark excitons that may be exploited for novel applications.

In this Letter, we identify a dark excitonic state (D) in monolayer MoS$_2$ that is blue detuned from the neutral exciton (X$^0$), and probe it using resonant SHG spectroscopy.
The cross section for SHG in 2D materials is enhanced by orders of magnitude when a two-photon resonance exists with bright excitonic states \cite{10.1038/nnano.2015.73,PhysRevLett.114.097403,10.1038/s41467-021-27213-8}.
The high quality of our sample allows us to clearly distinguish the SHG-exciton resonances arising from X$^0$ and various species of negatively charged trions, i.e., the triplet states (T$_{1,2,3}$) arising from hybridized inter- and intravalley trions \cite{PhysRevB.105.L041302,10.1039/D1NR03855A,10.1038/s41467-017-02286-6}.
Moreover, we observe SHG signatures of the dark exciton D, $\sim25$ meV blue detuned from X$^0$, via a pronounced anticrossing in the SHG signal as the excitation laser energy (2$\omega_p$) is tuned through the transition. 
This anticrossing is the result of coherent two-photon coupling in SHG enabled by the dark feature of D. 
We further strengthen our interpretation of this new dark state by probing its relaxation to localized excitons (LXs).
The relaxation is detected via two-photon PL-excitation spectroscopy \cite{10.1038/nature13734}, i.e., we monitor the spectral intensity of LX emission as the laser is tuned through the two-photon resonance with D.
The relaxation is found to be weak when the MoS$_2$ is placed on a planar substrate but is significantly enhanced for MoS$_2$ integrated into a nanobeam cavity.
This observation is fully consistent with the enhanced scattering between MoS$_2$ excitonic states due to cavity phononic modes \cite{PhysRevLett.130.126901}.
Our results demonstrate a coherent coupling between the dark exciton and $N=2$ photons and open the way to engineer nonlinear physics through structuring the nanoscale environment.


\begin{figure*}
    \includegraphics[width=\linewidth]{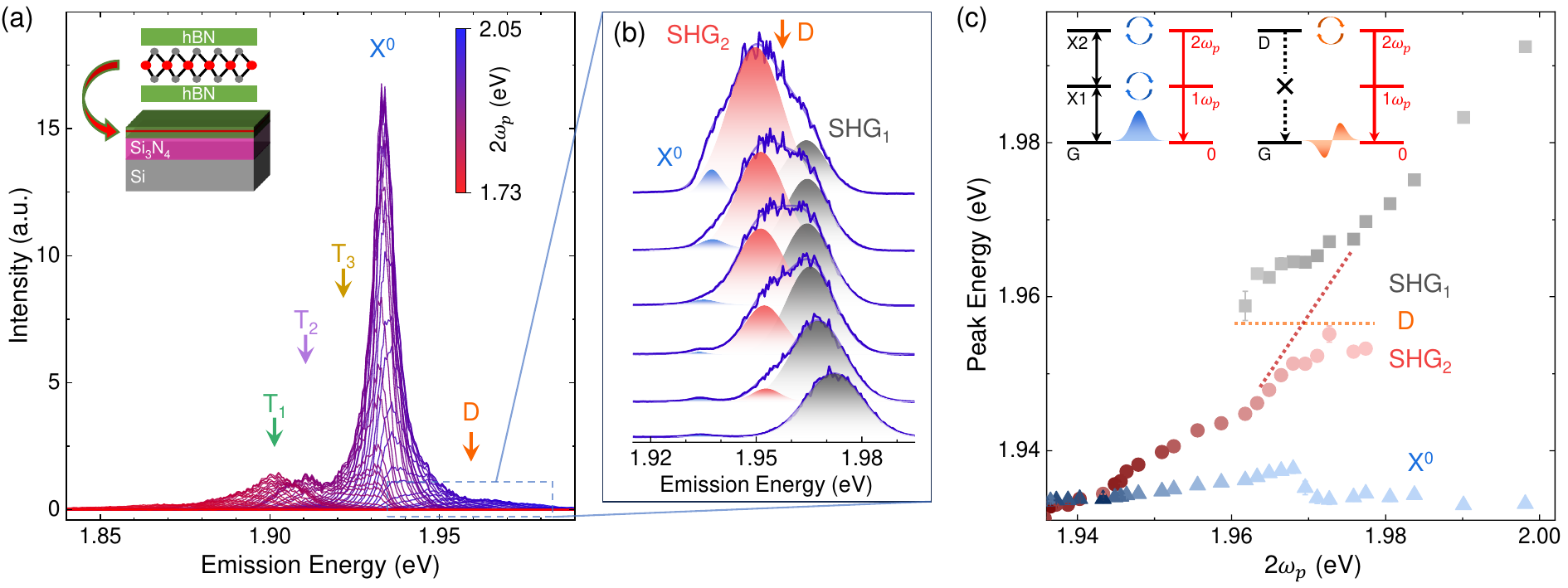}
    \caption{\label{f1}
        (a) Spectra for varying excitation laser energy $2\omega_p$.
        Intensity enhancement at SHG-exciton resonances is observed for bright excitons X$^0$ and T$_{1,2,3}$, while not for the dark state D.
        Inset is the schematic of the hBN/MoS$_2$/hBN heterostructure on the planar Si$_3$N$_4$/Si substrate.
        (b) Spectra near the SHG-D resonance. A splitting and anticrossing of the SHG signals (SHG$_1$ and SHG$_2$) is observed that reveals D at 25 meV blue detuned from X$^0$.
        (c) Peak energies extracted by fitting.
        The dotted lines denote bare energies without coupling.
        Insets are the schematic showing that the two-photon strong coupling for bright exciton denoted by X2 requires the mediation of another excitonic state denoted by X1.
        In contrast, the dark state D can directly and coherently couple to two photons, due to the opposite parity between two-photon and one-photon coupling.
    }
\end{figure*}

We begin by investigating SHG spectroscopy recorded from an hBN/MoS$_2$/hBN heterostructure on a planar Si$_3$N$_4$/Si substrate, as schematically depicted in the inset in Fig. \ref{f1}(a).
The monolayer MoS$_2$ is encapsulated by upper (15 nm) and lower (55 nm) hBN cladding layers, which suppresses the dielectric disorder and narrows the excitonic linewidths toward the homogeneous limit \cite{PhysRevX.7.021026,10.1038/s41598-017-09739-4}.
The sample is cooled down to 8 K and excited using a tunable pulsed laser with a spot size $\sim 1\ \mathrm{\mu m}$ and a CW equivalent excitation power of 15.2 mW.
The pulse length is $\sim$ 100 fs at a repetition frequency of 80 MHz.

We explore the resonant SHG signal as the two-photon energy of the excitation laser 2$\omega_p$ is tuned from 1.73 to 2.05 eV.
Typical data are presented in Fig. \ref{f1}(a).
A significant intensity enhancement is clearly observed with the peak emission energy at 1.933 eV, as the feature for the SHG-X$^0$ resonance \cite{PhysRevLett.114.097403,10.1038/nnano.2015.73,10.1038/s41467-021-27213-8}.
Besides X$^0$, we also observe the SHG resonance of the intravalley singlet trion T$_1$ at 1.902 eV, intervalley singlet trion T$_2$ at 1.911 eV, and intervalley triplet trion T$_3$ at 1.921 eV \cite{PhysRevB.105.L041302,10.1039/D1NR03855A,10.1038/s41467-017-02286-6}, as labeled in Fig. \ref{f1}(a).
Here, the SHG resonances for T$_{1,2,3}$ exhibit similar intensities, which is quite different to situations in PL spectroscopy (Fig. \ref{sf5} in Supplemental Material \cite{supplement}) where T$_1$ is much stronger than T$_{2,3}$.
We explain this difference between PL and SHG due to the fact that the resonant intensity enhancement of SHG is primarily determined by the oscillator strength of the respective transitions \cite{10.1038/nnano.2015.73}. 
The three trions T$_{1,2,3}$ in our sample have similar oscillator strength as predicted in theoretical calculations \cite{10.1063/5.0012971}, consistent with our observations.
In contrast, the relative intensity of T$_{1,2,3}$ observed in luminescence is determined by the interplay between their oscillator strength and population dynamics \cite{10.1038/nphys3604,PhysRevLett.130.126901,PhysRevB.105.L041302}, thereby differing from the resonances observed in SHG spectroscopy.

In Fig. \ref{f1}(b) we use an enlarged view to focus on a series of SHG spectra with the peak emission energy from $1.95-1.97$ eV.
Two distinct peaks labeled SHG$_1$ and SHG$_2$, highlighted by the red and gray shaded regions respectively, are clearly resolved in the data with a pronounced anticrossing centred at 1.958 eV.
At this point the two peaks have equal intensity and they are split by $\sim 13$ meV as shown in Fig. \ref{f1}(c).
The observed anticrossing is indicative of quantum interference in the strong light-matter interaction regime, but driven by the two-photon component of the driving field \cite{10.1038/s41567-018-0384-5,10.1038/s41467-021-21547-z,10.1088/0143-0807/37/2/025802}.
We attribute this resonance in the SHG intensity and the anticrossing to a new dark exciton state D, based on the observations that (i) it is not observed in PL or linear absorption spectroscopy, and (ii) a dipole forbidden dark state can coherently couple to the two photon components of the driving field \cite{PhysRevLett.130.083602,PhysRevLett.122.043601,PhysRevLett.72.1001}.
The strength of the one-photon light-matter interaction is $\propto\vert\langle X_f \vert \hat{r} \vert X_i\rangle\vert^2$, where $X_{f(i)}$ is the final (initial) state of the emitter \cite{fitzpatrick2015,PhysRevLett.130.083602}.
The position operator $\hat{r}$ represents one-photon coupling.
As depicted schematically by the blue wave function in the inset in Fig. \ref{f1}(c), excitonic states having primarily even parity have nonzero coupling and are, therefore, bright dipole allowed transitions.
In contrast, the two-photon coupling operator $\hat{r}^2$ has opposite parity to $\hat{r}$ in the one-photon coupling.
As a result, the bright exciton is dark in the two-photon coupling \cite{PhysRevLett.130.083602}.
Analogously, the one-photon dark state depicted by the orange wave function in the inset in Fig. \ref{f1}(c), becomes bright for two-photon coupling, since $\vert\langle D_f \vert \hat{r}^2 \vert D_i\rangle\vert^2$ is nonzero where $D_{f(i)}$ is the final (initial) state \cite{10.1038/nature13734}.

This difference of parity fundamentally determines the nature of the light-matter interaction.
Anticrossing arising from two-photon coupling has been reported for bright excitons in 2D materials \cite{10.1038/s41567-018-0384-5,10.1038/s41467-021-21547-z} and quantum dots \cite{PhysRevLett.120.213901}.
However, as depicted schematically in the inset in Fig. \ref{f1}(c), in these cases an intermediate excitonic state, denoted by X1 on the figure, exists close to the one-photon resonance.
X2 can transit to X1 by one photon emission, and thereby, the state X1 mediates two-photon coupling for X2.
The bright excitons X$^0$ and T$_{1,2,3}$ correspond to the situation when no intermediate state exists at their half energy around 0.96 eV \cite{10.1038/s41467-021-24102-y}.
Thus, while SHG spectroscopy has been widely reported for monolayer transition metal dichalcogenides, anticrossings indicative of strong two-photon coupling have not been observed.
In contrast, the dark exciton D directly couples to two photons without the participation of an intermediate state as schematically depicted in the inset in Fig. \ref{f1}(c).
As such, coherent two-phonon coupling is allowed giving rise to the anticrossing we observe in experiment.

\begin{figure}
    \includegraphics[width=\linewidth]{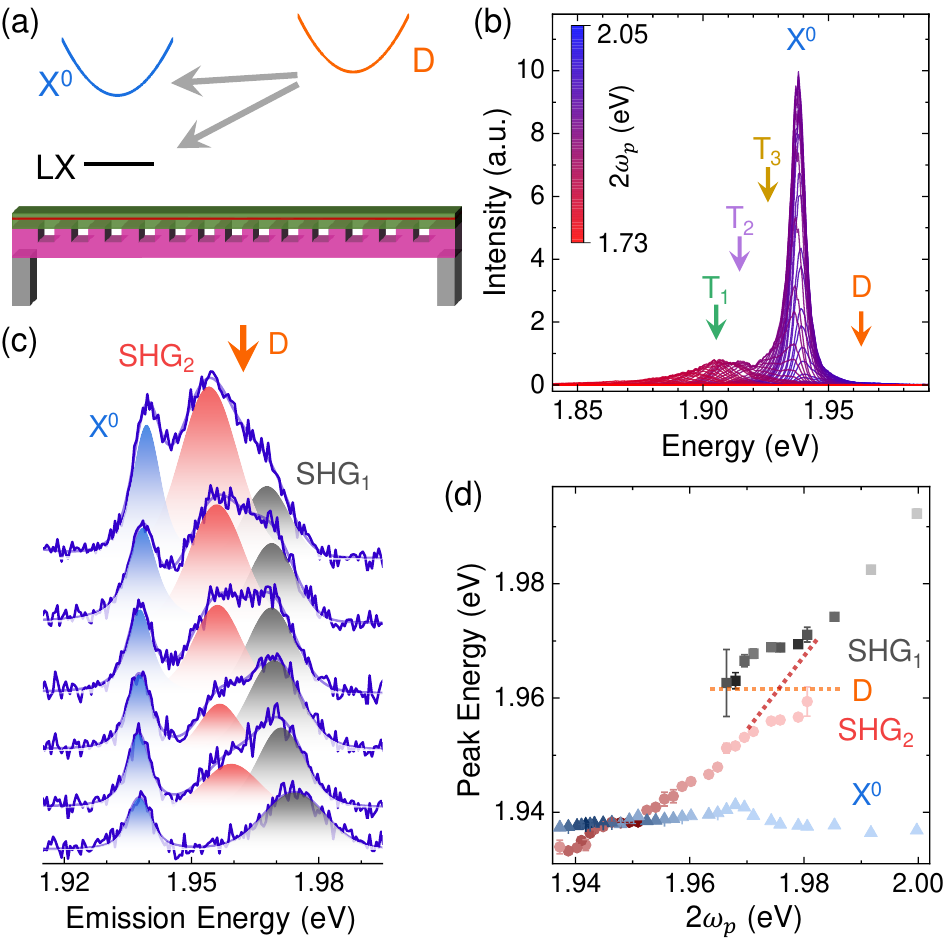}
    \caption{\label{f2}
        (a) Schematic of the hBN/MoS$_2$/hBN heterostructure integrated in the nanobeam cavity.
        Gray arrows in the upper panel depict that the intervalley scattering is enhanced in the cavity \cite{PhysRevLett.130.126901}.
        (b) The intensity enhancement at the resonances to bright excitons (X$^0$ and T$_{1,2,3}$) and (c) anticrossing at the resonance to D are observed, same to Figs. \ref{f1}(a) and \ref{f1}(b).
        (d) Peak energies extracted by fitting.
        Here D-X$^0$ energy detuning is still 25 meV.
        This strengthens that D arises from the excitonic feature of MoS$_2$.
    }
\end{figure}

The observed dark state D is $\sim$ 25 meV blue detuned from X$^0$, a characteristic energy that would be consistent with LA phonon scattering of the electron between the K and K' or Q valleys \cite{PhysRevB.85.115317}.
Therefore, we tentatively suggest that D may be an intervalley exciton \cite{PhysRevLett.115.176801,10.1038/s41699-017-0035-1,PhysRevB.98.020301}, as schematically depicted in Fig. \ref{f2}(a).
The intervalley state at energy higher than X$^0$ is supported by theoretical predictions using the ab initio GW-Bethe-Salpeter equation method reported by D. Qiu et al. \cite{PhysRevLett.115.176801}.
Recent work has also reported an intravalley dark exciton state at an energy $>10$ meV above X$^0$ \cite{10.1021/acs.jpclett.3c02431}.
As of yet, no consensus exists in the literature for the precise origin of all dark exciton states \cite{PhysRevLett.115.257403,10.1038/s41467-020-17608-4,PhysRevB.100.125413,10.1088/2053-1583/abb876,10.1038/ncomms14776,10.1088/2053-1583/aae953,PhysRevMaterials.2.014002}.
As such that the precise identification of D requires further investigation.
Nonetheless, our experimental observation of the dark exciton using resonant SHG spectroscopy is a robust and reproducible experimental observation, and the 25 meV blue detuning relative to X$^0$ is found to be a general result across multiple samples.

Our assignment of D as an intervalley state is further supported by SHG spectroscopy performed on the MoS$_2$ integrated in the high-Q nanobeam cavity.
The structure of our nanobeam cavity is schematically depicted in Fig.~\ref{f2}(a), and the fabrication and characterization details can be found in Refs. \cite{PhysRevLett.128.237403,PhysRevLett.130.126901}.
In the center region of the nanobeam, the photonic crystal periodicity (distance between nanoholes) is chirped to create the photonic and phononic band gap confinement of the cavity \cite{10.1038/nature08524}.
Recent work reported that the intervalley scattering between different excitonic states is significantly enhanced by cavity phononic modes \cite{PhysRevLett.130.126901}, because they provide intermediate replica states to enhance the exciton-phonon coupling.
As such, if D is an intervalley exciton, its incoherent relaxation to bright excitons will be greatly enhanced in the cavity, as schematically depicted by the gray arrows in Fig. \ref{f2}(a).
This prediction agrees remarkably with our experimental observations as we discuss next.

To illustrate this, in Fig. \ref{f2}(b) we present the spectra recorded from the cavity.
The intensity enhancement of the SHG-exciton resonances for X$^0$ at 1.937 eV and T$_{1,2,3}$ at 1.906, 1.914, and 1.925 eV is observed, similar to the case of the planar substrate in Fig. \ref{f1}(a).
Furthermore, as presented in the raw spectra in Fig. \ref{f2}(c) and fitting results in Fig. \ref{f2}(d), the anticrossing for the SHG-D resonance centered at 1.962 eV is clearly observed.
The energy differences between the excitons, particularly the 25 meV D-X$^0$ detuning, agree remarkably with those obtained in the planar substrate shown in Fig. \ref{f1}, supporting the overall reproducibility of the SHG spectroscopy and the excitonic feature of D arising from monolayer MoS$_2$.

\begin{figure}
    \includegraphics[width=\linewidth]{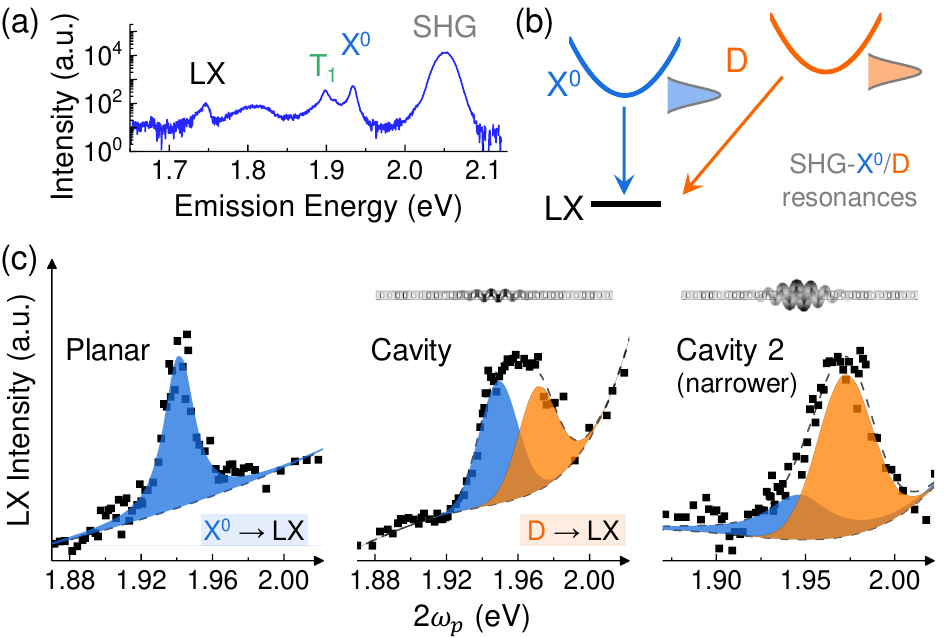}
    \caption{\label{f3}
        (a) One off-resonance spectrum showing the PL peak energy of X$^0$, T$_1$ and LX.
        (b) Arrows depict that LX peak is dominated by two-photon PL relaxed from higher energy states such as X$^0$ and D.
        Small peaks depict that LX intensity could be enhanced when the population X$^0$ and D are enhanced by SHG resonance.
        (c) LX intensity with the varying excitation energy $2\omega_p$ recorded in three cases.
        Blue and orange peaks correspond to the enhanced relaxation predicted in (b).
        D$\rightarrow$LX relaxation is suppressed on planar substrate while enabled in cavities, fully consistent to the scattering enhanced by cavity phononic environment as predicted in Fig. \ref{f2}(a).
        Insets depict that the narrower cavity has smaller bending rigidity and thus stronger phononic effects.
    }
\end{figure}

Moreover, we observe a significantly different relative intensity of the various spectral features when the MoS$_2$ is integrated in the cavity.
By comparing Figs. \ref{f1}(b) and \ref{f2}(c), we observe that the X$^0$ peak intensity relative to the SHG$_{1,2}$ features in the cavity is much larger than that on the planar substrate.
This is consistent with the prediction that the relaxation of D to bright excitons (D$\rightarrow$X$^0$) is enhanced in the cavity.
However, this relaxation is not very straightforward in experiment, because the X$^0$ peak arises from two contributions, including the two-photon PL (relaxation from higher energy states) and the SHG signal enhanced by the X$^0$ excitonic state \cite{10.1063/1.5100593}.
When the SHG is strongly detuned, the X$^0$ peak is dominated by the two-photon PL and exhibits the intrinsic emission peak energy that is independent from the excitation energy $2\omega_p$.
In contrast, when the SHG is near resonant, the X$^0$ peak is dominated by the enhanced SHG signal and the peak energy will shift with $2\omega_p$ \cite{10.1038/nnano.2015.73}.
The former corresponds to $2\omega_p > 1.98$ eV, and the latter corresponds to $2\omega_p < 1.97$ eV, respectively, as shown by the peak energy of X$^0$ in Figs. \ref{f1}(c) and \ref{f2}(d).
Around the SHG-D resonance both effects contribute, discussed in detail in Fig. \ref{sf3} in Supplemental Material \cite{supplement}.

Despite the multifaceted response for the X$^0$ peak, the peak of the localized excitons (LX) is always dominated by the two-photon PL relaxed from higher energy states such as X$^0$ or D \cite{10.1038/s41467-021-27213-8}.
LXs in MoS$_2$ originate from crystalline defects and emit in an energy band around $\sim 1.75$ eV as shown by the example in Fig. \ref{f3}(a).
When the SHG is near resonant to X$^0$ or D, the population of these free excitons as well as the corresponding relaxation to LX will be enhanced, as schematically depicted in Fig. \ref{f3}(b).
Therefore, resonant peaks observed in the two-photon PL-excitation spectroscopy of LX are proportional to the rates of relaxation from the higher energy free excitonic states.
The experimental results are presented in Fig. \ref{f3}(c), showing the LX intensity with varying $2\omega_p$ recorded from the planar substrate, the cavity, and another cavity in narrower nanobeam (details in Fig. \ref{sf1} in Supplemental Material \cite{supplement}), respectively.
Dashed lines in Fig. \ref{f3}(c) indicate a baseline emission, which gradually increases with $2\omega_p$, explained by above-gap absorption \cite{10.1038/s41467-017-02286-6}.
The enhanced LX emission, depicted by the blue shaded peak in Fig. \ref{f3}(c), centered at the SHG-X$^0$ resonance reveals the relaxation from X$^0$ into LX, and the orange shaded area centered at SHG-D resonance reveals the relaxation from D.
Clearly, when the monolayer MoS$_2$ is on the planar substrate, the LX intensity exhibits a single enhancement centered at the SHG-X$^0$ resonance, indicating that the D$\rightarrow$LX relaxation is suppressed.
In contrast, in cavities we observe a broadened feature indicating a double-peaked enhancement arising from both the SHG-X$^0$ and SHG-D resonance, indicating the D$\rightarrow$LX relaxation is enabled by the cavity.
These results are fully consistent with the cavity mediated intervalley scattering between different excitonic states \cite{PhysRevLett.130.126901}.
Comparing the effect of different cavities, the narrower nanobeam has smaller bending rigidity and therefore stronger phononic effects \cite{PhysRevLett.130.126901,2210.00150,10.1021/acs.nanolett.2c00613}.
This is fully consistent with the the stronger D$\rightarrow$LX relaxation in the narrower cavity shown in Fig. \ref{f3}(c).


In summary, we performed resonant SHG spectroscopy to probe a dark excitonic state D in monolayer MoS$_2$.
The dark exciton exhibits coherent two-photon strong coupling which manifests itself as an anticrossing observed in SHG spectroscopy. Our work shows that nonlinear optical spectroscopy methods are capable to provide new insights into excitonic photophysics, activating dipole forbidden transitions via coherent two-photon spectroscopy. 
Moreover, since dark excitons have unique features such as the antisymmetry and no spontaneous emission, their quantum interference with incident optical fields provide strong potential in nonlinear quantum devices based on 2D materials. 

\begin{acknowledgments}
    All authors gratefully acknowledge the German Science Foundation (DFG) for financial support via grants FI 947/8-1, DI 2013/5-1 and SPP-2244, as well as the clusters of excellence MCQST (EXS-2111) and e-conversion (EXS-2089).
    C. Q. and V. V. gratefully acknowledge the Alexander v. Humboldt foundation for financial support in the framework of their fellowship programme.
    We thank Paulo Eduardo de Faria Junior for his help in theoretical discussions.
\end{acknowledgments}

\input{refer.bbl}


\section*{Supplementary Material}
\setcounter{figure}{0}
\renewcommand{\thefigure}{S\arabic{figure}}
\setcounter{equation}{0}
\renewcommand{\theequation}{S\arabic{equation}}


\begin{figure*}
    \includegraphics[width=0.5\linewidth]{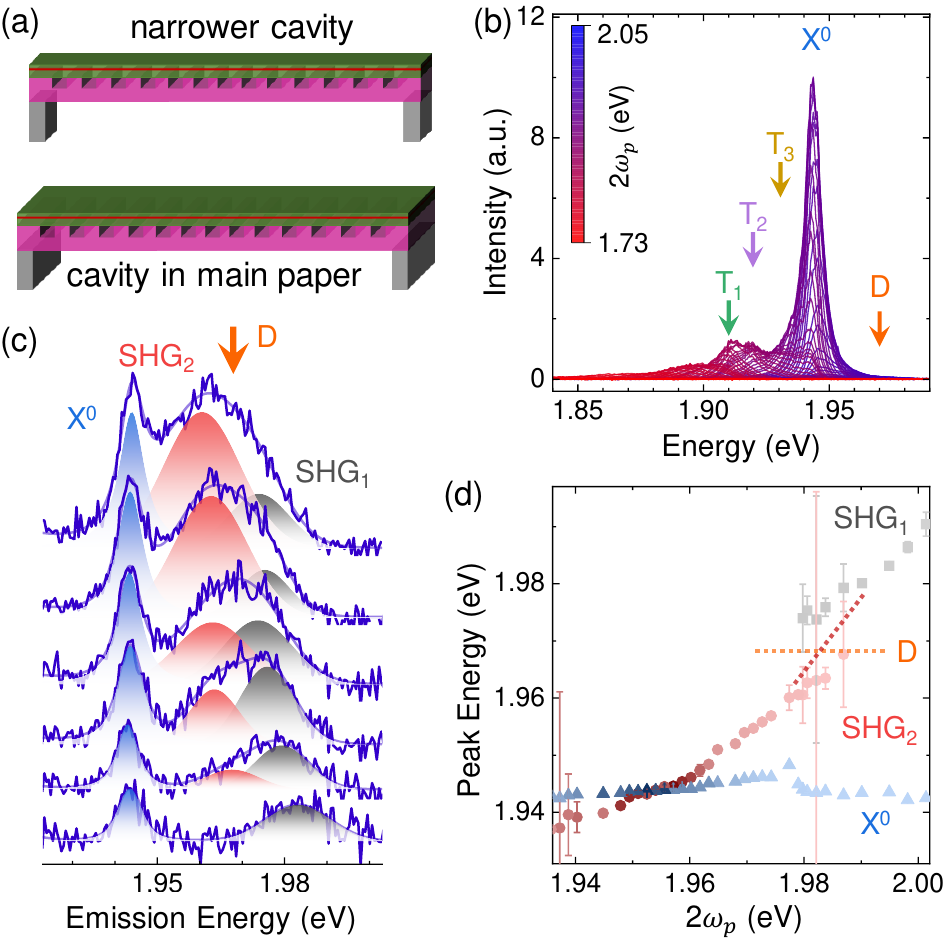}
    \caption{\label{sf1}
        Reproducibility strengthened in another narrower cavity.
        (a) Schematic depicts the different nanobeam width.
        This cavity has the nanobeam width of 420 nm, narrower than the cavity we discussed in the main paper (Fig. \ref{f2}) which has the width of 520 nm.
        (b) The intensity enhancement at the resonances to bright excitons (X$^0$ and T$_{1,2,3}$) and (c) anticrossing at the resonance to D are again observed.
        (d) Peak energies extracted by fitting.
        Here X$^0$ is at 1.944 eV, T$_{1,2,3}$ are at 1.912, 1.919 and 1.932 eV, and D is at 1.968 eV, respectively.
        D-X$^0$ energy detuning is 24 meV.
        These results further strengthens the conclusions in the main paper.
    }
\end{figure*}

\begin{figure*}
    \includegraphics[width=\linewidth]{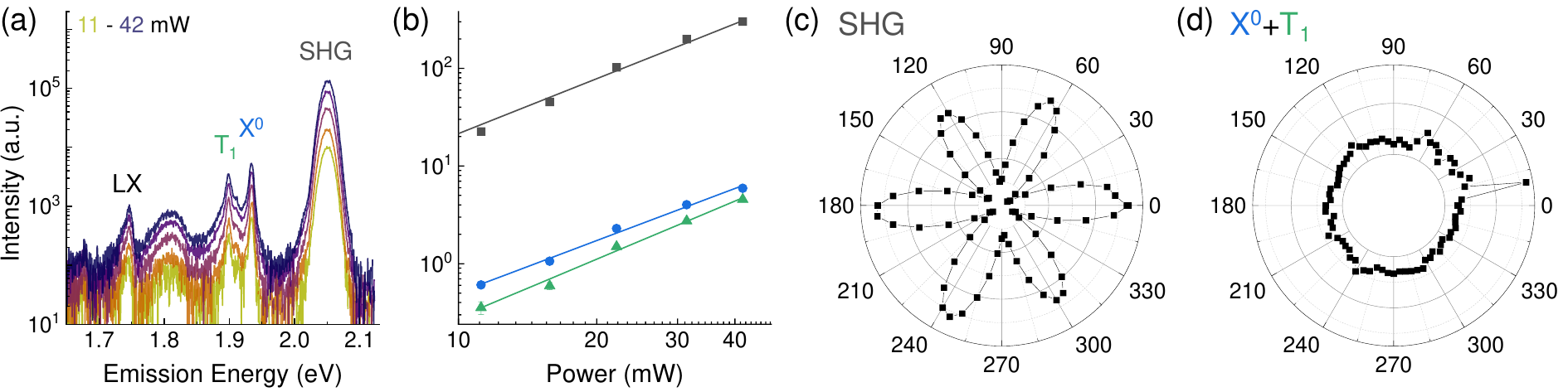}
    \caption{\label{sf2}
        Basic characterisation.
        (a) Power dependent spectra recorded with $2\omega_p$ of 2.05 eV (single photon $\omega_p$ at 1.02 eV) for CW equivalent excitation powers ranging from 11 to 42 mW.
        SHG peak and two-photon PL peaks arising from X$^0$, T$_1$ and LX are observed.
        (b) Peak intensities of SHG, X$^0$ and T$_1$. 
        All of them exhibit a slope $\sim$ 2 in the power dependence.
        (c) Intensity of SHG and (d) two-photon PL vs. the polarization angle of excitation laser.
        We note that in some cases, e.g., non-encapsulated monolayer TMDs, the X$^0$ peak has been reported to have a strong intensity more than one magnitude higher than the SHG peak, and exhibit clear polarization dependence indicating a mixture of enhanced SHG and two-photon PL.
        This can be explained by the fact that the scattering effects in non-encapsulated TMDs are strong and complex, resulting in the strong X$^0$ peak even under non-resonant excitation.
        However, this is not our case.
        As shown in (a)-(d), in our sample the intensity of X$^0$ peak is more than one order of magnitude smaller than the SHG, and little polarization dependence is observed.
        These results indicate that when the SHG is strongly detuned, the X$^0$ peak recorded from our sample is dominated by two-photon PL.
    }
\end{figure*}        
        
\begin{figure*}
    \includegraphics[width=0.66\linewidth]{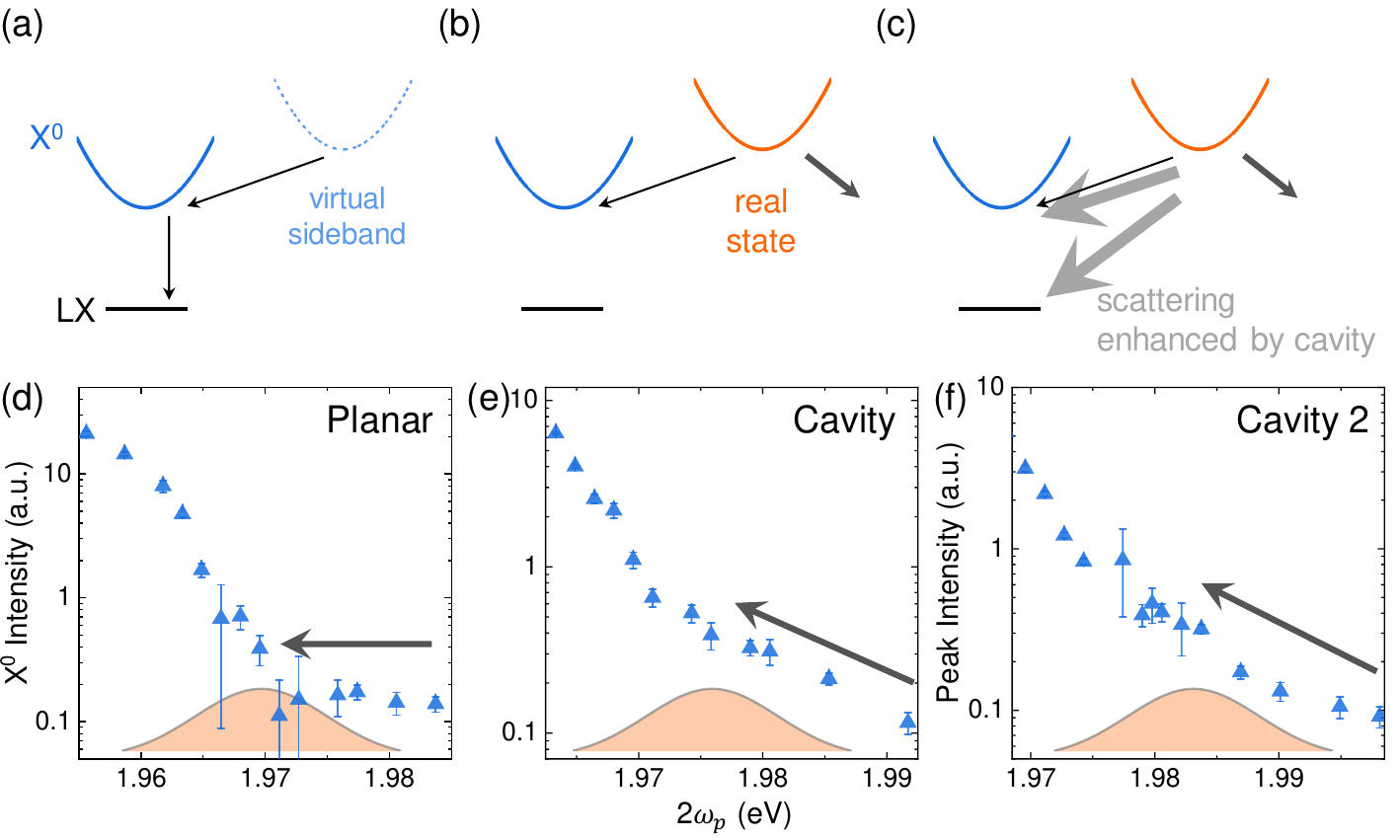}
    \caption{\label{sf3}
        Control of D$\rightarrow$X$^0$ relaxation, exhibiting same behavior of D$\rightarrow$LX relaxation in the main paper.
        (a) If D is a just a phonon-induced virtual sideband, relaxation of D to X$^0$ and LX should always be observed.
        But this is not the case in our experiment.
        (b) In the case that D is a real excitonic state, it has other relaxation approaches as schematically depicted by the bold arrow.
        This explains the observation that D$\rightarrow$LX relaxation is suppressed on the planar substrate.
        (c) In cavities, the intervalley scattering is enhanced.
        This explains that D$\rightarrow$LX relaxation is observed in cavities.
        (d)-(e) Although X$^0$ peak consists both enhanced SHG signals and two-photon PL emission thus is complex, we can still observe the D$\rightarrow$X$^0$ relaxation controlled by the cavity.
        At the right side of SHG-D resonance (denoted by the orange peak), the SHG-X$^0$ detuning is large, and thereby, we can expect the X$^0$ peak is dominated by two-photon PL including the contribution from D$\rightarrow$X$^0$ relaxation.
        As shown, this relaxation is (d) suppressed on planar substrate but (e)(f) enabled in cavities, consistent to the behavior of LX.
    }
\end{figure*}

\begin{figure*}
    \includegraphics[width=\linewidth]{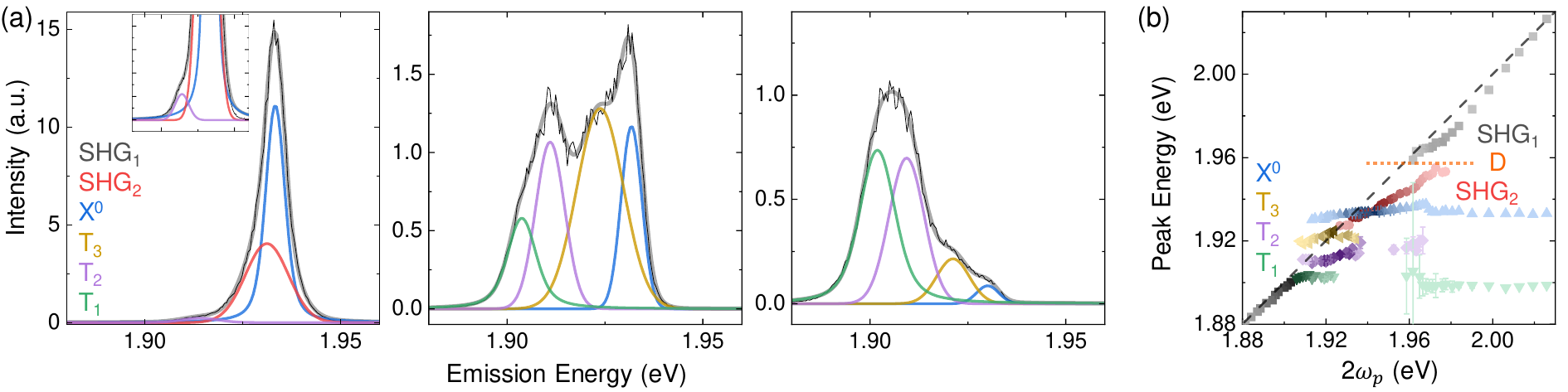}
    \caption{\label{sf4}
        Fitting method.
        (a) Fitting examples of the SHG spectroscopy recorded in the planar substrate.
        The intrinsic SHG from the MoS$_2$ lattice has the Gaussian lineshape.
        The exciton density of states have the Lorentz lineshape.
        As such, we use multiple Voigt peaks (convolution of Lorentz and Gaussian) to fit the spectra.
        When the SHG is tuning from the X$^0$ to T$_1$, the multiple peaks arising from fine excitonic structures overlap, which might introduce some fluctuations in the fitting result as shown in (b).
        Nonetheless, in this work we mainly focus on the SHG-D resonance at the energy above X$^0$, thus our conclusions are not affected.
    }
\end{figure*}

\begin{figure*}
    \includegraphics[width=\linewidth]{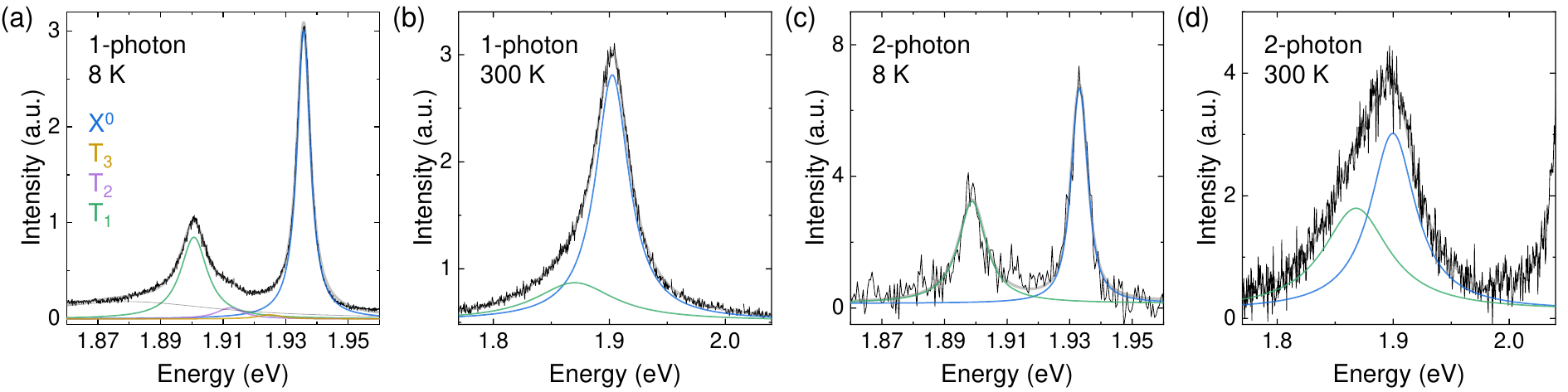}
    \caption{\label{sf5}
        PL peak of excitons.
        (a) One-photon (at 2.33 eV) excitation recorded at 8 K and (b) 300 K from a typical hBN-encapsulated MoS$_2$.
        (c) Two-photon (at 2.05 eV) excitation recorded at 8 K and 300 K from the bare case.
        The dark state D is not observed from the PL spectra.
        Trions T$_{2,3}$ can be distinguished at low temperature but have very weak intensities.
    }
\end{figure*}

\begin{figure*}
    \includegraphics[width=0.66\linewidth]{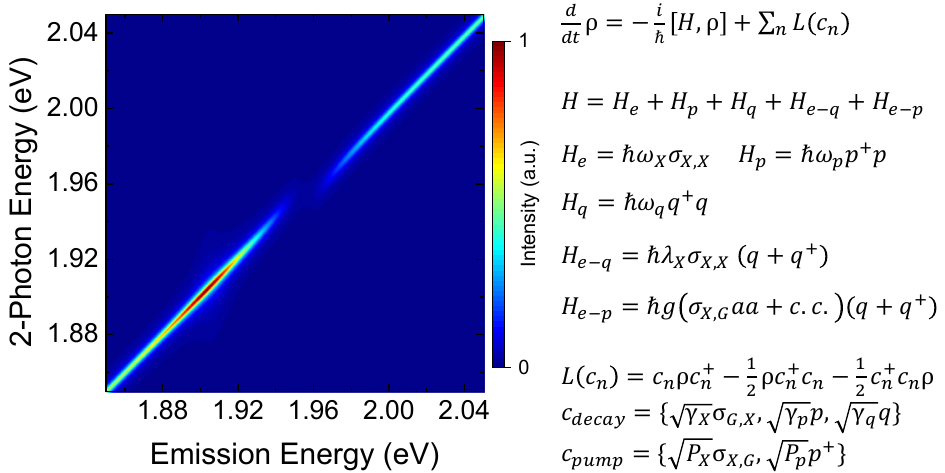}
    \caption{\label{sf6}
        Numerical calculation using the master equation theory.
        $\sigma_{A,B}=|A \rangle \langle B|$ ($A,B\in\{X,G\}$) is the Dirac operator for the exciton with energy $\omega_X$, $p^{+}$/$p$ are the ladder operators for photons with energy $\omega_p$, and $q^{+}$/$q$ are the ladder operators for phonons with energy $\omega_q$.
        $\lambda_{X}$ is the strength for the emitter-phonon coupling, and $g$ is the strength for the photon-polaron coupling.
        ${\cal L}(c_n)$ describes Markovian processes corresponding to the decay rate of exciton $\gamma_X$, decay rate of photon $\gamma_p$, decay rate of phonon $\gamma_q$, pump rate of exciton $P_X$, and pump rate of photon $P_p$, respectively.
        We use parameters $\omega_X=1930$, $\omega_q=25$, $\lambda_{X}=1$, $g=6$, $\gamma_X=5$, $\gamma_p=2$, $\gamma_q=10$, $P_X=0.02$, $P_p=0.1$, and varying excitation laser energy $\omega_p=900-1050$ to calculate the two-photon emission spectra shown in the map.
        The energy unit meV is omitted in the calculation.
        As shown, the calculation well reproduces the anticrossing in our experimental observations.
    }
\end{figure*}

\begin{figure*}
    \includegraphics[width=0.66\linewidth]{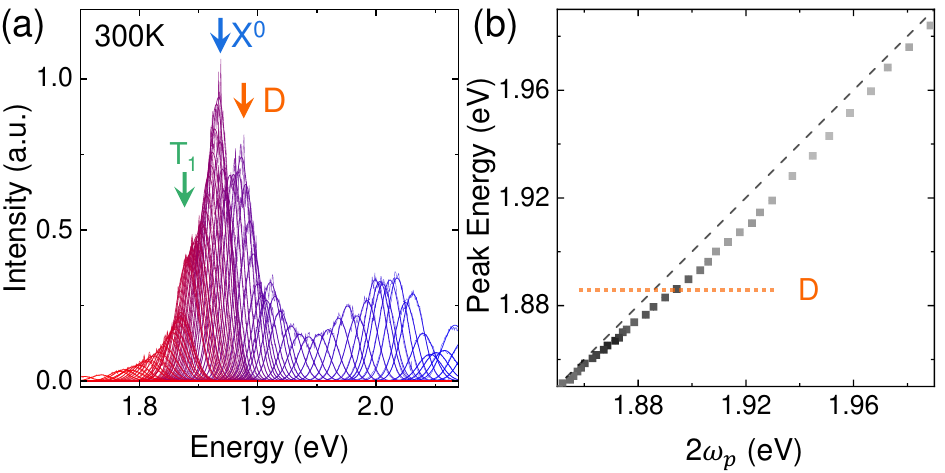}
    \caption{\label{sf7}
        Temperature dependence.
        (a) Resonant SHG spectroscopy of the bare heterostructure at room temperature 300 K.
        The intensity enhancement is observed for the resonance of SHG to X$^0$, T$_1$ and D as denoted.
        This indicates that at room temperature, the scattering between different excitonic states is significantly stronger than that at low temperature.
        This results in the mixture of dark and bright states, and thereby, D has a bit oscillator strength that enhances SHG.
        (b) Peak energies extracted by single peak fitting.
        D is $\sim$ 20 meV blue detuned to X$^0$, consistent with the data at low temperature.
        The peak energy of SHG is generally same as $2\omega_p$ when strongly detuned but smaller than $2\omega_p$ when resonant to excitons.
        This is explained by that near resonance the SHG intensity is inhomogeneously enhanced by the excitonic states along with complex dispersions in the energy level of 2D systems.
        Similar phenomena including the deviation and nonlinearity between SHG peak energy and $2\omega_p$ have been observed in other SHG spectroscopy of monolayer TMDs.
    }
\end{figure*}

\begin{figure*}
    \includegraphics[width=0.66\linewidth]{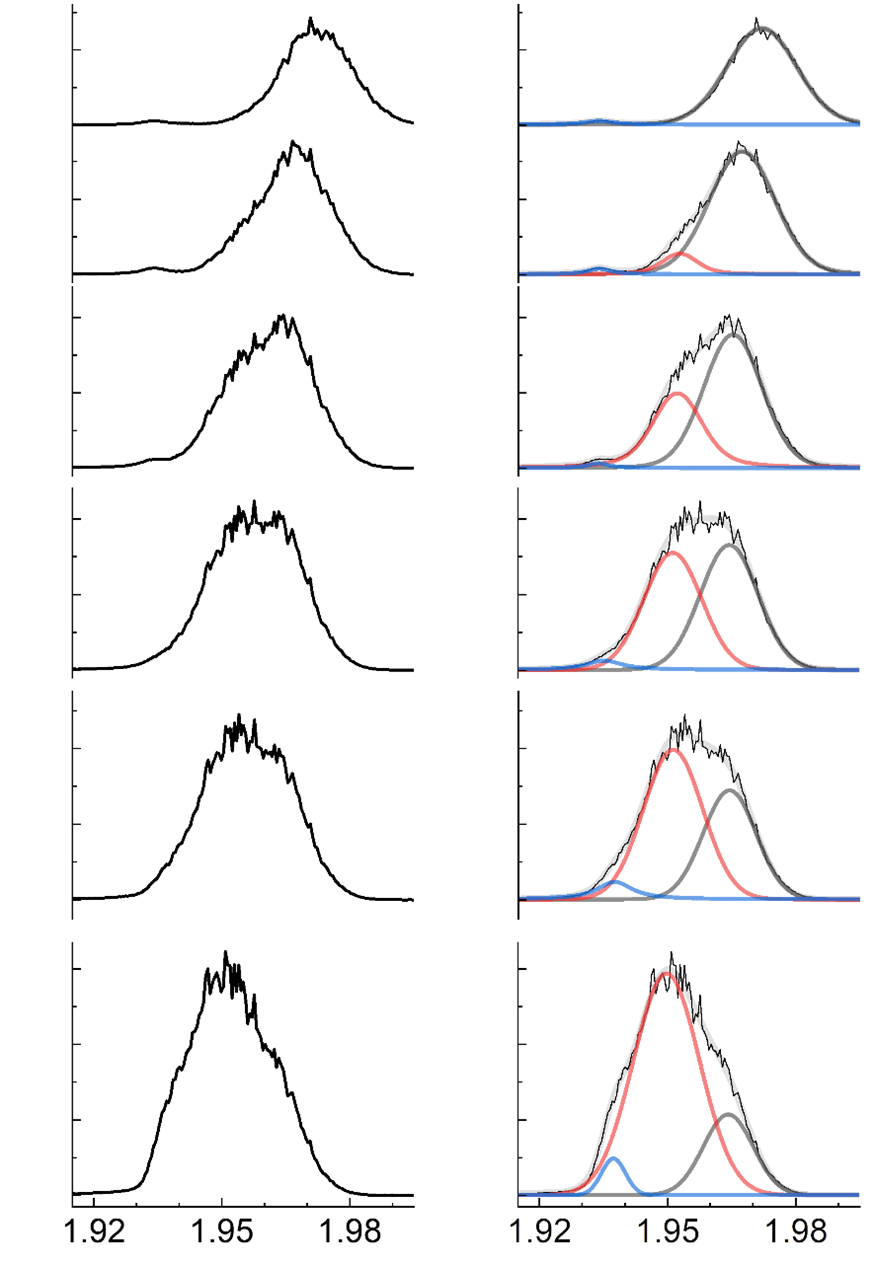}
    \caption{\label{sf8}
        The left panels show the detuning dependent raw spectra corresponding to the Fig. \ref{f1} in the main paper, and the right panels show the multi-peak fits. Two SHG peaks (gray and red) and the anticrossing are clearly distinguished even from the raw spectra.
    }
\end{figure*}

\begin{figure*}
    \includegraphics[width=0.66\linewidth]{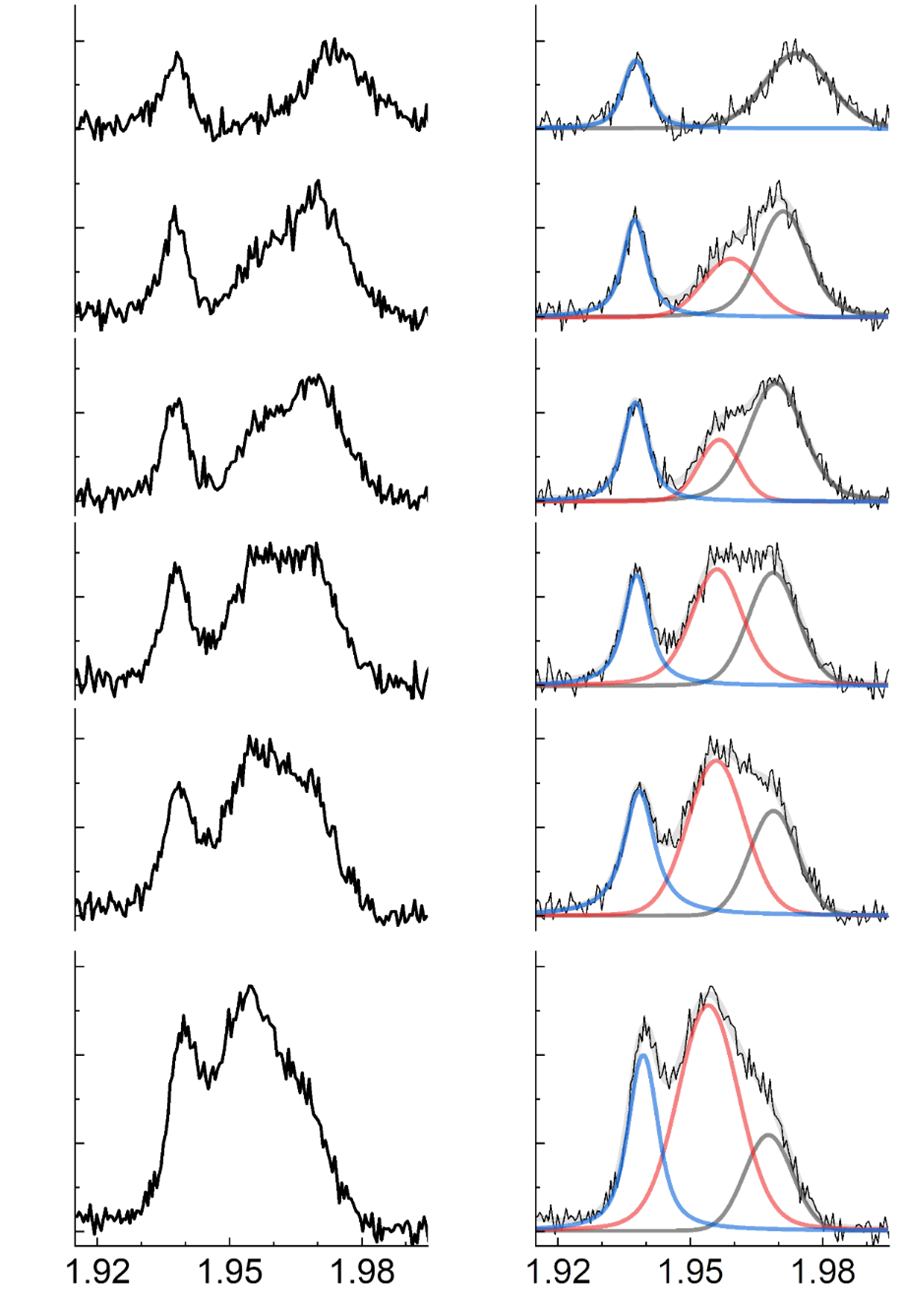}
    \caption{\label{sf9}
        The left panels show the detuning dependent raw spectra corresponding to the Fig. \ref{f2} in the main paper, and the right panels show the multi-peak fits. Two SHG peaks (gray and red) and the anticrossing are clearly distinguished even from the raw spectra.
    }
\end{figure*}

\end{document}

%% file: refer.bbl
%

%% file: main.bbl
\begin{thebibliography}{47}%
\makeatletter
\providecommand \@ifxundefined [1]{%
 \@ifx{#1\undefined}
}%
\providecommand \@ifnum [1]{%
 \ifnum #1\expandafter \@firstoftwo
 \else \expandafter \@secondoftwo
 \fi
}%
\providecommand \@ifx [1]{%
 \ifx #1\expandafter \@firstoftwo
 \else \expandafter \@secondoftwo
 \fi
}%
\providecommand \natexlab [1]{#1}%
\providecommand \enquote  [1]{``#1''}%
\providecommand \bibnamefont  [1]{#1}%
\providecommand \bibfnamefont [1]{#1}%
\providecommand \citenamefont [1]{#1}%
\providecommand \href@noop [0]{\@secondoftwo}%
\providecommand \href [0]{\begingroup \@sanitize@url \@href}%
\providecommand \@href[1]{\@@startlink{#1}\@@href}%
\providecommand \@@href[1]{\endgroup#1\@@endlink}%
\providecommand \@sanitize@url [0]{\catcode `\\12\catcode `\$12\catcode `\&12\catcode `\#12\catcode `\^12\catcode `\_12\catcode `\%12\relax}%
\providecommand \@@startlink[1]{}%
\providecommand \@@endlink[0]{}%
\providecommand \url  [0]{\begingroup\@sanitize@url \@url }%
\providecommand \@url [1]{\endgroup\@href {#1}{\urlprefix }}%
\providecommand \urlprefix  [0]{URL }%
\providecommand \Eprint [0]{\href }%
\providecommand \doibase [0]{https://doi.org/}%
\providecommand \selectlanguage [0]{\@gobble}%
\providecommand \bibinfo  [0]{\@secondoftwo}%
\providecommand \bibfield  [0]{\@secondoftwo}%
\providecommand \translation [1]{[#1]}%
\providecommand \BibitemOpen [0]{}%
\providecommand \bibitemStop [0]{}%
\providecommand \bibitemNoStop [0]{.\EOS\space}%
\providecommand \EOS [0]{\spacefactor3000\relax}%
\providecommand \BibitemShut  [1]{\csname bibitem#1\endcsname}%
\let\auto@bib@innerbib\@empty
\bibitem [{\citenamefont {Stegeman}\ and\ \citenamefont {Stegeman}(2012)}]{stegeman2012}%
  \BibitemOpen
  \bibfield  {author} {\bibinfo {author} {\bibfnamefont {G.}~\bibnamefont {Stegeman}}\ and\ \bibinfo {author} {\bibfnamefont {R.}~\bibnamefont {Stegeman}},\ }\href {https://www.wiley.com/en-us/Nonlinear+Optics%3A+Phenomena%2C+Materials+and+Devices-p-9781118268032} {\emph {\bibinfo {title} {Nonlinear Optics: Phenomena, Materials and Devices}}},\ Wiley Series in Pure and Applied Optics\ (\bibinfo  {publisher} {Wiley},\ \bibinfo {year} {2012})\BibitemShut {NoStop}%
\bibitem [{\citenamefont {Autere}\ \emph {et~al.}(2018)\citenamefont {Autere}, \citenamefont {Jussila}, \citenamefont {Dai}, \citenamefont {Wang}, \citenamefont {Lipsanen},\ and\ \citenamefont {Sun}}]{10.1002/adma.201705963}%
  \BibitemOpen
  \bibfield  {author} {\bibinfo {author} {\bibfnamefont {A.}~\bibnamefont {Autere}}, \bibinfo {author} {\bibfnamefont {H.}~\bibnamefont {Jussila}}, \bibinfo {author} {\bibfnamefont {Y.}~\bibnamefont {Dai}}, \bibinfo {author} {\bibfnamefont {Y.}~\bibnamefont {Wang}}, \bibinfo {author} {\bibfnamefont {H.}~\bibnamefont {Lipsanen}},\ and\ \bibinfo {author} {\bibfnamefont {Z.}~\bibnamefont {Sun}},\ }\bibfield  {title} {\bibinfo {title} {Nonlinear optics with {2D} layered materials},\ }\href {https://doi.org/https://doi.org/10.1002/adma.201705963} {\bibfield  {journal} {\bibinfo  {journal} {Adv. Mater.}\ }\textbf {\bibinfo {volume} {30}},\ \bibinfo {pages} {1705963} (\bibinfo {year} {2018})}\BibitemShut {NoStop}%
\bibitem [{\citenamefont {Wen}\ \emph {et~al.}(2019)\citenamefont {Wen}, \citenamefont {Gong},\ and\ \citenamefont {Li}}]{doi.org/10.1002/inf2.12024}%
  \BibitemOpen
  \bibfield  {author} {\bibinfo {author} {\bibfnamefont {X.}~\bibnamefont {Wen}}, \bibinfo {author} {\bibfnamefont {Z.}~\bibnamefont {Gong}},\ and\ \bibinfo {author} {\bibfnamefont {D.}~\bibnamefont {Li}},\ }\bibfield  {title} {\bibinfo {title} {Nonlinear optics of two-dimensional transition metal dichalcogenides},\ }\href {https://doi.org/https://doi.org/10.1002/inf2.12024} {\bibfield  {journal} {\bibinfo  {journal} {InfoMat}\ }\textbf {\bibinfo {volume} {1}},\ \bibinfo {pages} {317} (\bibinfo {year} {2019})}\BibitemShut {NoStop}%
\bibitem [{\citenamefont {Gu}\ \emph {et~al.}(2012)\citenamefont {Gu}, \citenamefont {Petrone}, \citenamefont {McMillan}, \citenamefont {van~der Zande}, \citenamefont {Yu}, \citenamefont {Lo}, \citenamefont {Kwong}, \citenamefont {Hone},\ and\ \citenamefont {Wong}}]{10.1038/nphoton.2012.147}%
  \BibitemOpen
  \bibfield  {author} {\bibinfo {author} {\bibfnamefont {T.}~\bibnamefont {Gu}}, \bibinfo {author} {\bibfnamefont {N.}~\bibnamefont {Petrone}}, \bibinfo {author} {\bibfnamefont {J.~F.}\ \bibnamefont {McMillan}}, \bibinfo {author} {\bibfnamefont {A.}~\bibnamefont {van~der Zande}}, \bibinfo {author} {\bibfnamefont {M.}~\bibnamefont {Yu}}, \bibinfo {author} {\bibfnamefont {G.~Q.}\ \bibnamefont {Lo}}, \bibinfo {author} {\bibfnamefont {D.~L.}\ \bibnamefont {Kwong}}, \bibinfo {author} {\bibfnamefont {J.}~\bibnamefont {Hone}},\ and\ \bibinfo {author} {\bibfnamefont {C.~W.}\ \bibnamefont {Wong}},\ }\bibfield  {title} {\bibinfo {title} {Regenerative oscillation and four-wave mixing in graphene optoelectronics},\ }\href {https://doi.org/10.1038/nphoton.2012.147} {\bibfield  {journal} {\bibinfo  {journal} {Nat. Photonics}\ }\textbf {\bibinfo {volume} {6}},\ \bibinfo {pages} {554} (\bibinfo {year} {2012})}\BibitemShut {NoStop}%
\bibitem [{\citenamefont {Seyler}\ \emph {et~al.}(2015)\citenamefont {Seyler}, \citenamefont {Schaibley}, \citenamefont {Gong}, \citenamefont {Rivera}, \citenamefont {Jones}, \citenamefont {Wu}, \citenamefont {Yan}, \citenamefont {Mandrus}, \citenamefont {Yao},\ and\ \citenamefont {Xu}}]{10.1038/nnano.2015.73}%
  \BibitemOpen
  \bibfield  {author} {\bibinfo {author} {\bibfnamefont {K.~L.}\ \bibnamefont {Seyler}}, \bibinfo {author} {\bibfnamefont {J.~R.}\ \bibnamefont {Schaibley}}, \bibinfo {author} {\bibfnamefont {P.}~\bibnamefont {Gong}}, \bibinfo {author} {\bibfnamefont {P.}~\bibnamefont {Rivera}}, \bibinfo {author} {\bibfnamefont {A.~M.}\ \bibnamefont {Jones}}, \bibinfo {author} {\bibfnamefont {S.}~\bibnamefont {Wu}}, \bibinfo {author} {\bibfnamefont {J.}~\bibnamefont {Yan}}, \bibinfo {author} {\bibfnamefont {D.~G.}\ \bibnamefont {Mandrus}}, \bibinfo {author} {\bibfnamefont {W.}~\bibnamefont {Yao}},\ and\ \bibinfo {author} {\bibfnamefont {X.}~\bibnamefont {Xu}},\ }\bibfield  {title} {\bibinfo {title} {Electrical control of second-harmonic generation in a {WSe2} monolayer transistor},\ }\href {https://doi.org/10.1038/nnano.2015.73} {\bibfield  {journal} {\bibinfo  {journal} {Nat. Nanotechnol.}\ }\textbf {\bibinfo {volume} {10}},\ \bibinfo {pages} {407} (\bibinfo {year} {2015})}\BibitemShut {NoStop}%
\bibitem [{\citenamefont {Wang}\ \emph {et~al.}(2015)\citenamefont {Wang}, \citenamefont {Marie}, \citenamefont {Gerber}, \citenamefont {Amand}, \citenamefont {Lagarde}, \citenamefont {Bouet}, \citenamefont {Vidal}, \citenamefont {Balocchi},\ and\ \citenamefont {Urbaszek}}]{PhysRevLett.114.097403}%
  \BibitemOpen
  \bibfield  {author} {\bibinfo {author} {\bibfnamefont {G.}~\bibnamefont {Wang}}, \bibinfo {author} {\bibfnamefont {X.}~\bibnamefont {Marie}}, \bibinfo {author} {\bibfnamefont {I.}~\bibnamefont {Gerber}}, \bibinfo {author} {\bibfnamefont {T.}~\bibnamefont {Amand}}, \bibinfo {author} {\bibfnamefont {D.}~\bibnamefont {Lagarde}}, \bibinfo {author} {\bibfnamefont {L.}~\bibnamefont {Bouet}}, \bibinfo {author} {\bibfnamefont {M.}~\bibnamefont {Vidal}}, \bibinfo {author} {\bibfnamefont {A.}~\bibnamefont {Balocchi}},\ and\ \bibinfo {author} {\bibfnamefont {B.}~\bibnamefont {Urbaszek}},\ }\bibfield  {title} {\bibinfo {title} {Giant enhancement of the optical second-harmonic emission of {${\mathrm{WSe}}_{2}$} monolayers by laser excitation at exciton resonances},\ }\href {https://doi.org/10.1103/PhysRevLett.114.097403} {\bibfield  {journal} {\bibinfo  {journal} {Phys. Rev. Lett.}\ }\textbf {\bibinfo {volume} {114}},\ \bibinfo {pages} {097403} (\bibinfo {year} {2015})}\BibitemShut {NoStop}%
\bibitem [{\citenamefont {Shree}\ \emph {et~al.}(2021)\citenamefont {Shree}, \citenamefont {Lagarde}, \citenamefont {Lombez}, \citenamefont {Robert}, \citenamefont {Balocchi}, \citenamefont {Watanabe}, \citenamefont {Taniguchi}, \citenamefont {Marie}, \citenamefont {Gerber}, \citenamefont {Glazov}, \citenamefont {Golub}, \citenamefont {Urbaszek},\ and\ \citenamefont {Paradisanos}}]{10.1038/s41467-021-27213-8}%
  \BibitemOpen
  \bibfield  {author} {\bibinfo {author} {\bibfnamefont {S.}~\bibnamefont {Shree}}, \bibinfo {author} {\bibfnamefont {D.}~\bibnamefont {Lagarde}}, \bibinfo {author} {\bibfnamefont {L.}~\bibnamefont {Lombez}}, \bibinfo {author} {\bibfnamefont {C.}~\bibnamefont {Robert}}, \bibinfo {author} {\bibfnamefont {A.}~\bibnamefont {Balocchi}}, \bibinfo {author} {\bibfnamefont {K.}~\bibnamefont {Watanabe}}, \bibinfo {author} {\bibfnamefont {T.}~\bibnamefont {Taniguchi}}, \bibinfo {author} {\bibfnamefont {X.}~\bibnamefont {Marie}}, \bibinfo {author} {\bibfnamefont {I.~C.}\ \bibnamefont {Gerber}}, \bibinfo {author} {\bibfnamefont {M.~M.}\ \bibnamefont {Glazov}}, \bibinfo {author} {\bibfnamefont {L.~E.}\ \bibnamefont {Golub}}, \bibinfo {author} {\bibfnamefont {B.}~\bibnamefont {Urbaszek}},\ and\ \bibinfo {author} {\bibfnamefont {I.}~\bibnamefont {Paradisanos}},\ }\bibfield  {title} {\bibinfo {title} {Interlayer exciton mediated second harmonic generation in bilayer {MoS2}},\ }\href {https://doi.org/10.1038/s41467-021-27213-8} {\bibfield  {journal} {\bibinfo  {journal} {Nat. Commun.}\ }\textbf {\bibinfo {volume} {12}},\ \bibinfo {pages} {6894} (\bibinfo {year} {2021})}\BibitemShut {NoStop}%
\bibitem [{\citenamefont {Klimmer}\ \emph {et~al.}(2021)\citenamefont {Klimmer}, \citenamefont {Ghaebi}, \citenamefont {Gan}, \citenamefont {George}, \citenamefont {Turchanin}, \citenamefont {Cerullo},\ and\ \citenamefont {Soavi}}]{10.1038/s41566-021-00859-y}%
  \BibitemOpen
  \bibfield  {author} {\bibinfo {author} {\bibfnamefont {S.}~\bibnamefont {Klimmer}}, \bibinfo {author} {\bibfnamefont {O.}~\bibnamefont {Ghaebi}}, \bibinfo {author} {\bibfnamefont {Z.}~\bibnamefont {Gan}}, \bibinfo {author} {\bibfnamefont {A.}~\bibnamefont {George}}, \bibinfo {author} {\bibfnamefont {A.}~\bibnamefont {Turchanin}}, \bibinfo {author} {\bibfnamefont {G.}~\bibnamefont {Cerullo}},\ and\ \bibinfo {author} {\bibfnamefont {G.}~\bibnamefont {Soavi}},\ }\bibfield  {title} {\bibinfo {title} {All-optical polarization and amplitude modulation of second-harmonic generation in atomically thin semiconductors},\ }\href {https://doi.org/10.1038/s41566-021-00859-y} {\bibfield  {journal} {\bibinfo  {journal} {Nat. Photonics}\ }\textbf {\bibinfo {volume} {15}},\ \bibinfo {pages} {837} (\bibinfo {year} {2021})}\BibitemShut {NoStop}%
\bibitem [{\citenamefont {Paradisanos}\ \emph {et~al.}(2022)\citenamefont {Paradisanos}, \citenamefont {Raven}, \citenamefont {Amand}, \citenamefont {Robert}, \citenamefont {Renucci}, \citenamefont {Watanabe}, \citenamefont {Taniguchi}, \citenamefont {Gerber}, \citenamefont {Marie},\ and\ \citenamefont {Urbaszek}}]{PhysRevB.105.115420}%
  \BibitemOpen
  \bibfield  {author} {\bibinfo {author} {\bibfnamefont {I.}~\bibnamefont {Paradisanos}}, \bibinfo {author} {\bibfnamefont {A.~M.~S.}\ \bibnamefont {Raven}}, \bibinfo {author} {\bibfnamefont {T.}~\bibnamefont {Amand}}, \bibinfo {author} {\bibfnamefont {C.}~\bibnamefont {Robert}}, \bibinfo {author} {\bibfnamefont {P.}~\bibnamefont {Renucci}}, \bibinfo {author} {\bibfnamefont {K.}~\bibnamefont {Watanabe}}, \bibinfo {author} {\bibfnamefont {T.}~\bibnamefont {Taniguchi}}, \bibinfo {author} {\bibfnamefont {I.~C.}\ \bibnamefont {Gerber}}, \bibinfo {author} {\bibfnamefont {X.}~\bibnamefont {Marie}},\ and\ \bibinfo {author} {\bibfnamefont {B.}~\bibnamefont {Urbaszek}},\ }\bibfield  {title} {\bibinfo {title} {Second harmonic generation control in twisted bilayers of transition metal dichalcogenides},\ }\href {https://doi.org/10.1103/PhysRevB.105.115420} {\bibfield  {journal} {\bibinfo  {journal} {Phys. Rev. B}\ }\textbf {\bibinfo {volume} {105}},\ \bibinfo {pages} {115420} (\bibinfo {year} {2022})}\BibitemShut {NoStop}%
\bibitem [{\citenamefont {Dogadov}\ \emph {et~al.}(2022)\citenamefont {Dogadov}, \citenamefont {Trovatello}, \citenamefont {Yao}, \citenamefont {Soavi},\ and\ \citenamefont {Cerullo}}]{10.1002/lpor.202100726}%
  \BibitemOpen
  \bibfield  {author} {\bibinfo {author} {\bibfnamefont {O.}~\bibnamefont {Dogadov}}, \bibinfo {author} {\bibfnamefont {C.}~\bibnamefont {Trovatello}}, \bibinfo {author} {\bibfnamefont {B.}~\bibnamefont {Yao}}, \bibinfo {author} {\bibfnamefont {G.}~\bibnamefont {Soavi}},\ and\ \bibinfo {author} {\bibfnamefont {G.}~\bibnamefont {Cerullo}},\ }\bibfield  {title} {\bibinfo {title} {Parametric nonlinear optics with layered materials and related heterostructures},\ }\href {https://doi.org/https://doi.org/10.1002/lpor.202100726} {\bibfield  {journal} {\bibinfo  {journal} {Laser Photonics Rev.}\ }\textbf {\bibinfo {volume} {16}},\ \bibinfo {pages} {2100726} (\bibinfo {year} {2022})}\BibitemShut {NoStop}%
\bibitem [{\citenamefont {Villafa\~ne}\ \emph {et~al.}(2023)\citenamefont {Villafa\~ne}, \citenamefont {Scaparra}, \citenamefont {Rieger}, \citenamefont {Appel}, \citenamefont {Trivedi}, \citenamefont {Zhu}, \citenamefont {Jarman}, \citenamefont {Oliver}, \citenamefont {Taylor}, \citenamefont {Finley},\ and\ \citenamefont {M\"uller}}]{PhysRevLett.130.083602}%
  \BibitemOpen
  \bibfield  {author} {\bibinfo {author} {\bibfnamefont {V.}~\bibnamefont {Villafa\~ne}}, \bibinfo {author} {\bibfnamefont {B.}~\bibnamefont {Scaparra}}, \bibinfo {author} {\bibfnamefont {M.}~\bibnamefont {Rieger}}, \bibinfo {author} {\bibfnamefont {S.}~\bibnamefont {Appel}}, \bibinfo {author} {\bibfnamefont {R.}~\bibnamefont {Trivedi}}, \bibinfo {author} {\bibfnamefont {T.}~\bibnamefont {Zhu}}, \bibinfo {author} {\bibfnamefont {J.}~\bibnamefont {Jarman}}, \bibinfo {author} {\bibfnamefont {R.~A.}\ \bibnamefont {Oliver}}, \bibinfo {author} {\bibfnamefont {R.~A.}\ \bibnamefont {Taylor}}, \bibinfo {author} {\bibfnamefont {J.~J.}\ \bibnamefont {Finley}},\ and\ \bibinfo {author} {\bibfnamefont {K.}~\bibnamefont {M\"uller}},\ }\bibfield  {title} {\bibinfo {title} {Three-photon excitation of {InGaN} quantum dots},\ }\href {https://doi.org/10.1103/PhysRevLett.130.083602} {\bibfield  {journal} {\bibinfo  {journal} {Phys. Rev. Lett.}\ }\textbf {\bibinfo {volume} {130}},\ \bibinfo {pages} {083602} (\bibinfo {year} {2023})}\BibitemShut {NoStop}%
\bibitem [{\citenamefont {Malekakhlagh}\ and\ \citenamefont {Rodriguez}(2019)}]{PhysRevLett.122.043601}%
  \BibitemOpen
  \bibfield  {author} {\bibinfo {author} {\bibfnamefont {M.}~\bibnamefont {Malekakhlagh}}\ and\ \bibinfo {author} {\bibfnamefont {A.~W.}\ \bibnamefont {Rodriguez}},\ }\bibfield  {title} {\bibinfo {title} {Quantum {Rabi} model with two-photon relaxation},\ }\href {https://doi.org/10.1103/PhysRevLett.122.043601} {\bibfield  {journal} {\bibinfo  {journal} {Phys. Rev. Lett.}\ }\textbf {\bibinfo {volume} {122}},\ \bibinfo {pages} {043601} (\bibinfo {year} {2019})}\BibitemShut {NoStop}%
\bibitem [{\citenamefont {Fr\"ohlich}\ \emph {et~al.}(1994)\citenamefont {Fr\"ohlich}, \citenamefont {Itoh},\ and\ \citenamefont {Pahlke-Lerch}}]{PhysRevLett.72.1001}%
  \BibitemOpen
  \bibfield  {author} {\bibinfo {author} {\bibfnamefont {D.}~\bibnamefont {Fr\"ohlich}}, \bibinfo {author} {\bibfnamefont {M.}~\bibnamefont {Itoh}},\ and\ \bibinfo {author} {\bibfnamefont {C.}~\bibnamefont {Pahlke-Lerch}},\ }\bibfield  {title} {\bibinfo {title} {Two-photon spectroscopy of odd-parity states},\ }\href {https://doi.org/10.1103/PhysRevLett.72.1001} {\bibfield  {journal} {\bibinfo  {journal} {Phys. Rev. Lett.}\ }\textbf {\bibinfo {volume} {72}},\ \bibinfo {pages} {1001} (\bibinfo {year} {1994})}\BibitemShut {NoStop}%
\bibitem [{\citenamefont {Ye}\ \emph {et~al.}(2014)\citenamefont {Ye}, \citenamefont {Cao}, \citenamefont {O'Brien}, \citenamefont {Zhu}, \citenamefont {Yin}, \citenamefont {Wang}, \citenamefont {Louie},\ and\ \citenamefont {Zhang}}]{10.1038/nature13734}%
  \BibitemOpen
  \bibfield  {author} {\bibinfo {author} {\bibfnamefont {Z.}~\bibnamefont {Ye}}, \bibinfo {author} {\bibfnamefont {T.}~\bibnamefont {Cao}}, \bibinfo {author} {\bibfnamefont {K.}~\bibnamefont {O'Brien}}, \bibinfo {author} {\bibfnamefont {H.}~\bibnamefont {Zhu}}, \bibinfo {author} {\bibfnamefont {X.}~\bibnamefont {Yin}}, \bibinfo {author} {\bibfnamefont {Y.}~\bibnamefont {Wang}}, \bibinfo {author} {\bibfnamefont {S.~G.}\ \bibnamefont {Louie}},\ and\ \bibinfo {author} {\bibfnamefont {X.}~\bibnamefont {Zhang}},\ }\bibfield  {title} {\bibinfo {title} {Probing excitonic dark states in single-layer tungsten disulphide},\ }\href {https://doi.org/10.1038/nature13734} {\bibfield  {journal} {\bibinfo  {journal} {Nature}\ }\textbf {\bibinfo {volume} {513}},\ \bibinfo {pages} {214} (\bibinfo {year} {2014})}\BibitemShut {NoStop}%
\bibitem [{\citenamefont {Slavcheva}\ and\ \citenamefont {Kavokin}(2013)}]{PhysRevB.88.085321}%
  \BibitemOpen
  \bibfield  {author} {\bibinfo {author} {\bibfnamefont {G.}~\bibnamefont {Slavcheva}}\ and\ \bibinfo {author} {\bibfnamefont {A.~V.}\ \bibnamefont {Kavokin}},\ }\bibfield  {title} {\bibinfo {title} {Polarization selection rules in exciton-based terahertz lasers},\ }\href {https://doi.org/10.1103/PhysRevB.88.085321} {\bibfield  {journal} {\bibinfo  {journal} {Phys. Rev. B}\ }\textbf {\bibinfo {volume} {88}},\ \bibinfo {pages} {085321} (\bibinfo {year} {2013})}\BibitemShut {NoStop}%
\bibitem [{\citenamefont {Lin}\ \emph {et~al.}(2019)\citenamefont {Lin}, \citenamefont {Bange},\ and\ \citenamefont {Lupton}}]{10.1038/s41567-018-0384-5}%
  \BibitemOpen
  \bibfield  {author} {\bibinfo {author} {\bibfnamefont {K.-Q.}\ \bibnamefont {Lin}}, \bibinfo {author} {\bibfnamefont {S.}~\bibnamefont {Bange}},\ and\ \bibinfo {author} {\bibfnamefont {J.~M.}\ \bibnamefont {Lupton}},\ }\bibfield  {title} {\bibinfo {title} {Quantum interference in second-harmonic generation from monolayer {WSe2}},\ }\href {https://doi.org/10.1038/s41567-018-0384-5} {\bibfield  {journal} {\bibinfo  {journal} {Nat. Phys.}\ }\textbf {\bibinfo {volume} {15}},\ \bibinfo {pages} {242} (\bibinfo {year} {2019})}\BibitemShut {NoStop}%
\bibitem [{\citenamefont {Qian}\ \emph {et~al.}(2018)\citenamefont {Qian}, \citenamefont {Wu}, \citenamefont {Song}, \citenamefont {Peng}, \citenamefont {Xie}, \citenamefont {Yang}, \citenamefont {Xiao}, \citenamefont {Steer}, \citenamefont {Thayne}, \citenamefont {Tang}, \citenamefont {Zuo}, \citenamefont {Jin}, \citenamefont {Gu},\ and\ \citenamefont {Xu}}]{PhysRevLett.120.213901}%
  \BibitemOpen
  \bibfield  {author} {\bibinfo {author} {\bibfnamefont {C.}~\bibnamefont {Qian}}, \bibinfo {author} {\bibfnamefont {S.}~\bibnamefont {Wu}}, \bibinfo {author} {\bibfnamefont {F.}~\bibnamefont {Song}}, \bibinfo {author} {\bibfnamefont {K.}~\bibnamefont {Peng}}, \bibinfo {author} {\bibfnamefont {X.}~\bibnamefont {Xie}}, \bibinfo {author} {\bibfnamefont {J.}~\bibnamefont {Yang}}, \bibinfo {author} {\bibfnamefont {S.}~\bibnamefont {Xiao}}, \bibinfo {author} {\bibfnamefont {M.~J.}\ \bibnamefont {Steer}}, \bibinfo {author} {\bibfnamefont {I.~G.}\ \bibnamefont {Thayne}}, \bibinfo {author} {\bibfnamefont {C.}~\bibnamefont {Tang}}, \bibinfo {author} {\bibfnamefont {Z.}~\bibnamefont {Zuo}}, \bibinfo {author} {\bibfnamefont {K.}~\bibnamefont {Jin}}, \bibinfo {author} {\bibfnamefont {C.}~\bibnamefont {Gu}},\ and\ \bibinfo {author} {\bibfnamefont {X.}~\bibnamefont {Xu}},\ }\bibfield  {title} {\bibinfo {title} {Two-photon {Rabi} splitting in a coupled system of a nanocavity and exciton complexes},\ }\href {https://doi.org/10.1103/PhysRevLett.120.213901} {\bibfield  {journal} {\bibinfo  {journal} {Phys. Rev. Lett.}\ }\textbf {\bibinfo {volume} {120}},\ \bibinfo {pages} {213901} (\bibinfo {year} {2018})}\BibitemShut {NoStop}%
\bibitem [{\citenamefont {Lin}\ \emph {et~al.}(2021)\citenamefont {Lin}, \citenamefont {Faria~Junior}, \citenamefont {Bauer}, \citenamefont {Peng}, \citenamefont {Monserrat}, \citenamefont {Gmitra}, \citenamefont {Fabian}, \citenamefont {Bange},\ and\ \citenamefont {Lupton}}]{10.1038/s41467-021-21547-z}%
  \BibitemOpen
  \bibfield  {author} {\bibinfo {author} {\bibfnamefont {K.-Q.}\ \bibnamefont {Lin}}, \bibinfo {author} {\bibfnamefont {P.~E.}\ \bibnamefont {Faria~Junior}}, \bibinfo {author} {\bibfnamefont {J.~M.}\ \bibnamefont {Bauer}}, \bibinfo {author} {\bibfnamefont {B.}~\bibnamefont {Peng}}, \bibinfo {author} {\bibfnamefont {B.}~\bibnamefont {Monserrat}}, \bibinfo {author} {\bibfnamefont {M.}~\bibnamefont {Gmitra}}, \bibinfo {author} {\bibfnamefont {J.}~\bibnamefont {Fabian}}, \bibinfo {author} {\bibfnamefont {S.}~\bibnamefont {Bange}},\ and\ \bibinfo {author} {\bibfnamefont {J.~M.}\ \bibnamefont {Lupton}},\ }\bibfield  {title} {\bibinfo {title} {Twist-angle engineering of excitonic quantum interference and optical nonlinearities in stacked {2D} semiconductors},\ }\href {https://doi.org/10.1038/s41467-021-21547-z} {\bibfield  {journal} {\bibinfo  {journal} {Nat. Commun.}\ }\textbf {\bibinfo {volume} {12}},\ \bibinfo {pages} {1553} (\bibinfo {year} {2021})}\BibitemShut {NoStop}%
\bibitem [{\citenamefont {Petri\'{c}}\ \emph {et~al.}(2023)\citenamefont {Petri\'{c}}, \citenamefont {Villafa\~ne}, \citenamefont {Qian}, \citenamefont {Kremser}, \citenamefont {Finley},\ and\ \citenamefont {M\"{u}ller}}]{10.1364/CLEO_FS.2023.FW4N.4}%
  \BibitemOpen
  \bibfield  {author} {\bibinfo {author} {\bibfnamefont {M.~M.}\ \bibnamefont {Petri\'{c}}}, \bibinfo {author} {\bibfnamefont {V.}~\bibnamefont {Villafa\~ne}}, \bibinfo {author} {\bibfnamefont {C.}~\bibnamefont {Qian}}, \bibinfo {author} {\bibfnamefont {M.}~\bibnamefont {Kremser}}, \bibinfo {author} {\bibfnamefont {J.~J.}\ \bibnamefont {Finley}},\ and\ \bibinfo {author} {\bibfnamefont {K.}~\bibnamefont {M\"{u}ller}},\ }\bibfield  {title} {\bibinfo {title} {Quantum interference with interlayer excitons in moir\'{e} {MoSe2} homobilayers},\ }in\ \href {https://doi.org/10.1364/CLEO_FS.2023.FW4N.4} {\emph {\bibinfo {booktitle} {CLEO 2023}}}\ (\bibinfo  {publisher} {Optica Publishing Group},\ \bibinfo {year} {2023})\ p.\ \bibinfo {pages} {FW4N.4}\BibitemShut {NoStop}%
\bibitem [{\citenamefont {Klein}\ \emph {et~al.}(2022)\citenamefont {Klein}, \citenamefont {Florian}, \citenamefont {H\"otger}, \citenamefont {Steinhoff}, \citenamefont {Delhomme}, \citenamefont {Taniguchi}, \citenamefont {Watanabe}, \citenamefont {Jahnke}, \citenamefont {Holleitner}, \citenamefont {Potemski}, \citenamefont {Faugeras}, \citenamefont {Stier},\ and\ \citenamefont {Finley}}]{PhysRevB.105.L041302}%
  \BibitemOpen
  \bibfield  {author} {\bibinfo {author} {\bibfnamefont {J.}~\bibnamefont {Klein}}, \bibinfo {author} {\bibfnamefont {M.}~\bibnamefont {Florian}}, \bibinfo {author} {\bibfnamefont {A.}~\bibnamefont {H\"otger}}, \bibinfo {author} {\bibfnamefont {A.}~\bibnamefont {Steinhoff}}, \bibinfo {author} {\bibfnamefont {A.}~\bibnamefont {Delhomme}}, \bibinfo {author} {\bibfnamefont {T.}~\bibnamefont {Taniguchi}}, \bibinfo {author} {\bibfnamefont {K.}~\bibnamefont {Watanabe}}, \bibinfo {author} {\bibfnamefont {F.}~\bibnamefont {Jahnke}}, \bibinfo {author} {\bibfnamefont {A.~W.}\ \bibnamefont {Holleitner}}, \bibinfo {author} {\bibfnamefont {M.}~\bibnamefont {Potemski}}, \bibinfo {author} {\bibfnamefont {C.}~\bibnamefont {Faugeras}}, \bibinfo {author} {\bibfnamefont {A.~V.}\ \bibnamefont {Stier}},\ and\ \bibinfo {author} {\bibfnamefont {J.~J.}\ \bibnamefont {Finley}},\ }\bibfield  {title} {\bibinfo {title} {Trions in {${\mathrm{MoS}}_{2}$} are quantum superpositions of intra- and intervalley spin states},\ }\href {https://doi.org/10.1103/PhysRevB.105.L041302} {\bibfield  {journal} {\bibinfo  {journal} {Phys. Rev. B}\ }\textbf {\bibinfo {volume} {105}},\ \bibinfo {pages} {L041302} (\bibinfo {year} {2022})}\BibitemShut {NoStop}%
\bibitem [{\citenamefont {Grzeszczyk}\ \emph {et~al.}(2021)\citenamefont {Grzeszczyk}, \citenamefont {Olkowska-Pucko}, \citenamefont {Nogajewski}, \citenamefont {Watanabe}, \citenamefont {Taniguchi}, \citenamefont {Kossacki}, \citenamefont {Babiński},\ and\ \citenamefont {Molas}}]{10.1039/D1NR03855A}%
  \BibitemOpen
  \bibfield  {author} {\bibinfo {author} {\bibfnamefont {M.}~\bibnamefont {Grzeszczyk}}, \bibinfo {author} {\bibfnamefont {K.}~\bibnamefont {Olkowska-Pucko}}, \bibinfo {author} {\bibfnamefont {K.}~\bibnamefont {Nogajewski}}, \bibinfo {author} {\bibfnamefont {K.}~\bibnamefont {Watanabe}}, \bibinfo {author} {\bibfnamefont {T.}~\bibnamefont {Taniguchi}}, \bibinfo {author} {\bibfnamefont {P.}~\bibnamefont {Kossacki}}, \bibinfo {author} {\bibfnamefont {A.}~\bibnamefont {Babiński}},\ and\ \bibinfo {author} {\bibfnamefont {M.~R.}\ \bibnamefont {Molas}},\ }\bibfield  {title} {\bibinfo {title} {Exposing the trion{'}s fine structure by controlling the carrier concentration in hbn-encapsulated {MoS2}},\ }\href {https://doi.org/10.1039/D1NR03855A} {\bibfield  {journal} {\bibinfo  {journal} {Nanoscale}\ }\textbf {\bibinfo {volume} {13}},\ \bibinfo {pages} {18726} (\bibinfo {year} {2021})}\BibitemShut {NoStop}%
\bibitem [{\citenamefont {Dr{\"u}ppel}\ \emph {et~al.}(2017)\citenamefont {Dr{\"u}ppel}, \citenamefont {Deilmann}, \citenamefont {Kr{\"u}ger},\ and\ \citenamefont {Rohlfing}}]{10.1038/s41467-017-02286-6}%
  \BibitemOpen
  \bibfield  {author} {\bibinfo {author} {\bibfnamefont {M.}~\bibnamefont {Dr{\"u}ppel}}, \bibinfo {author} {\bibfnamefont {T.}~\bibnamefont {Deilmann}}, \bibinfo {author} {\bibfnamefont {P.}~\bibnamefont {Kr{\"u}ger}},\ and\ \bibinfo {author} {\bibfnamefont {M.}~\bibnamefont {Rohlfing}},\ }\bibfield  {title} {\bibinfo {title} {Diversity of trion states and substrate effects in the optical properties of an {MoS2} monolayer},\ }\href {https://doi.org/10.1038/s41467-017-02286-6} {\bibfield  {journal} {\bibinfo  {journal} {Nat. Commun.}\ }\textbf {\bibinfo {volume} {8}},\ \bibinfo {pages} {2117} (\bibinfo {year} {2017})}\BibitemShut {NoStop}%
\bibitem [{\citenamefont {Qian}\ \emph {et~al.}(2023)\citenamefont {Qian}, \citenamefont {Villafa\~ne}, \citenamefont {Petri\ifmmode~\acute{c}\else \'{c}\fi{}}, \citenamefont {Soubelet}, \citenamefont {Stier},\ and\ \citenamefont {Finley}}]{PhysRevLett.130.126901}%
  \BibitemOpen
  \bibfield  {author} {\bibinfo {author} {\bibfnamefont {C.}~\bibnamefont {Qian}}, \bibinfo {author} {\bibfnamefont {V.}~\bibnamefont {Villafa\~ne}}, \bibinfo {author} {\bibfnamefont {M.~M.}\ \bibnamefont {Petri\ifmmode~\acute{c}\else \'{c}\fi{}}}, \bibinfo {author} {\bibfnamefont {P.}~\bibnamefont {Soubelet}}, \bibinfo {author} {\bibfnamefont {A.~V.}\ \bibnamefont {Stier}},\ and\ \bibinfo {author} {\bibfnamefont {J.~J.}\ \bibnamefont {Finley}},\ }\bibfield  {title} {\bibinfo {title} {Coupling of {${\mathrm{MoS}}_{2}$} excitons with lattice phonons and cavity vibrational phonons in hybrid nanobeam cavities},\ }\href {https://doi.org/10.1103/PhysRevLett.130.126901} {\bibfield  {journal} {\bibinfo  {journal} {Phys. Rev. Lett.}\ }\textbf {\bibinfo {volume} {130}},\ \bibinfo {pages} {126901} (\bibinfo {year} {2023})}\BibitemShut {NoStop}%
\bibitem [{\citenamefont {Cadiz}\ \emph {et~al.}(2017)\citenamefont {Cadiz}, \citenamefont {Courtade}, \citenamefont {Robert}, \citenamefont {Wang}, \citenamefont {Shen}, \citenamefont {Cai}, \citenamefont {Taniguchi}, \citenamefont {Watanabe}, \citenamefont {Carrere}, \citenamefont {Lagarde}, \citenamefont {Manca}, \citenamefont {Amand}, \citenamefont {Renucci}, \citenamefont {Tongay}, \citenamefont {Marie},\ and\ \citenamefont {Urbaszek}}]{PhysRevX.7.021026}%
  \BibitemOpen
  \bibfield  {author} {\bibinfo {author} {\bibfnamefont {F.}~\bibnamefont {Cadiz}}, \bibinfo {author} {\bibfnamefont {E.}~\bibnamefont {Courtade}}, \bibinfo {author} {\bibfnamefont {C.}~\bibnamefont {Robert}}, \bibinfo {author} {\bibfnamefont {G.}~\bibnamefont {Wang}}, \bibinfo {author} {\bibfnamefont {Y.}~\bibnamefont {Shen}}, \bibinfo {author} {\bibfnamefont {H.}~\bibnamefont {Cai}}, \bibinfo {author} {\bibfnamefont {T.}~\bibnamefont {Taniguchi}}, \bibinfo {author} {\bibfnamefont {K.}~\bibnamefont {Watanabe}}, \bibinfo {author} {\bibfnamefont {H.}~\bibnamefont {Carrere}}, \bibinfo {author} {\bibfnamefont {D.}~\bibnamefont {Lagarde}}, \bibinfo {author} {\bibfnamefont {M.}~\bibnamefont {Manca}}, \bibinfo {author} {\bibfnamefont {T.}~\bibnamefont {Amand}}, \bibinfo {author} {\bibfnamefont {P.}~\bibnamefont {Renucci}}, \bibinfo {author} {\bibfnamefont {S.}~\bibnamefont {Tongay}}, \bibinfo {author} {\bibfnamefont {X.}~\bibnamefont {Marie}},\ and\ \bibinfo {author} {\bibfnamefont {B.}~\bibnamefont {Urbaszek}},\ }\bibfield  {title} {\bibinfo {title} {Excitonic linewidth approaching the homogeneous limit in {${\mathrm{MoS}}_{2}$}-based van der {Waals} heterostructures},\ }\href {https://doi.org/10.1103/PhysRevX.7.021026} {\bibfield  {journal} {\bibinfo  {journal} {Phys. Rev. X}\ }\textbf {\bibinfo {volume} {7}},\ \bibinfo {pages} {021026} (\bibinfo {year} {2017})}\BibitemShut {NoStop}%
\bibitem [{\citenamefont {Wierzbowski}\ \emph {et~al.}(2017)\citenamefont {Wierzbowski}, \citenamefont {Klein}, \citenamefont {Sigger}, \citenamefont {Straubinger}, \citenamefont {Kremser}, \citenamefont {Taniguchi}, \citenamefont {Watanabe}, \citenamefont {Wurstbauer}, \citenamefont {Holleitner}, \citenamefont {Kaniber}, \citenamefont {M{\"u}ller},\ and\ \citenamefont {Finley}}]{10.1038/s41598-017-09739-4}%
  \BibitemOpen
  \bibfield  {author} {\bibinfo {author} {\bibfnamefont {J.}~\bibnamefont {Wierzbowski}}, \bibinfo {author} {\bibfnamefont {J.}~\bibnamefont {Klein}}, \bibinfo {author} {\bibfnamefont {F.}~\bibnamefont {Sigger}}, \bibinfo {author} {\bibfnamefont {C.}~\bibnamefont {Straubinger}}, \bibinfo {author} {\bibfnamefont {M.}~\bibnamefont {Kremser}}, \bibinfo {author} {\bibfnamefont {T.}~\bibnamefont {Taniguchi}}, \bibinfo {author} {\bibfnamefont {K.}~\bibnamefont {Watanabe}}, \bibinfo {author} {\bibfnamefont {U.}~\bibnamefont {Wurstbauer}}, \bibinfo {author} {\bibfnamefont {A.~W.}\ \bibnamefont {Holleitner}}, \bibinfo {author} {\bibfnamefont {M.}~\bibnamefont {Kaniber}}, \bibinfo {author} {\bibfnamefont {K.}~\bibnamefont {M{\"u}ller}},\ and\ \bibinfo {author} {\bibfnamefont {J.~J.}\ \bibnamefont {Finley}},\ }\bibfield  {title} {\bibinfo {title} {Direct exciton emission from atomically thin transition metal dichalcogenide heterostructures near the lifetime limit},\ }\href {https://doi.org/10.1038/s41598-017-09739-4} {\bibfield  {journal} {\bibinfo  {journal} {Sci. Rep.}\ }\textbf {\bibinfo {volume} {7}},\ \bibinfo {pages} {12383} (\bibinfo {year} {2017})}\BibitemShut {NoStop}%
\bibitem []{supplement}%
  \BibitemOpen
  See Supplemental Material for the methods, the theoretical calculation, and the additional experimental data to strengthen the reproducibility
  \BibitemShut {NoStop}%
\bibitem [{\citenamefont {Zhumagulov}\ \emph {et~al.}(2020)\citenamefont {Zhumagulov}, \citenamefont {Vagov}, \citenamefont {Gulevich}, \citenamefont {Faria~Junior},\ and\ \citenamefont {Perebeinos}}]{10.1063/5.0012971}%
  \BibitemOpen
  \bibfield  {author} {\bibinfo {author} {\bibfnamefont {Y.~V.}\ \bibnamefont {Zhumagulov}}, \bibinfo {author} {\bibfnamefont {A.}~\bibnamefont {Vagov}}, \bibinfo {author} {\bibfnamefont {D.~R.}\ \bibnamefont {Gulevich}}, \bibinfo {author} {\bibfnamefont {P.~E.}\ \bibnamefont {Faria~Junior}},\ and\ \bibinfo {author} {\bibfnamefont {V.}~\bibnamefont {Perebeinos}},\ }\bibfield  {title} {\bibinfo {title} {Trion induced photoluminescence of a doped {MoS2} monolayer},\ }\bibfield  {journal} {\bibinfo  {journal} {J. Chem. Phys.}\ }\textbf {\bibinfo {volume} {153}},\ \href {https://doi.org/10.1063/5.0012971} {10.1063/5.0012971} (\bibinfo {year} {2020})\BibitemShut {NoStop}%
\bibitem [{\citenamefont {Jones}\ \emph {et~al.}(2016)\citenamefont {Jones}, \citenamefont {Yu}, \citenamefont {Schaibley}, \citenamefont {Yan}, \citenamefont {Mandrus}, \citenamefont {Taniguchi}, \citenamefont {Watanabe}, \citenamefont {Dery}, \citenamefont {Yao},\ and\ \citenamefont {Xu}}]{10.1038/nphys3604}%
  \BibitemOpen
  \bibfield  {author} {\bibinfo {author} {\bibfnamefont {A.~M.}\ \bibnamefont {Jones}}, \bibinfo {author} {\bibfnamefont {H.}~\bibnamefont {Yu}}, \bibinfo {author} {\bibfnamefont {J.~R.}\ \bibnamefont {Schaibley}}, \bibinfo {author} {\bibfnamefont {J.}~\bibnamefont {Yan}}, \bibinfo {author} {\bibfnamefont {D.~G.}\ \bibnamefont {Mandrus}}, \bibinfo {author} {\bibfnamefont {T.}~\bibnamefont {Taniguchi}}, \bibinfo {author} {\bibfnamefont {K.}~\bibnamefont {Watanabe}}, \bibinfo {author} {\bibfnamefont {H.}~\bibnamefont {Dery}}, \bibinfo {author} {\bibfnamefont {W.}~\bibnamefont {Yao}},\ and\ \bibinfo {author} {\bibfnamefont {X.}~\bibnamefont {Xu}},\ }\bibfield  {title} {\bibinfo {title} {Excitonic luminescence upconversion in a two-dimensional semiconductor},\ }\href {https://doi.org/10.1038/nphys3604} {\bibfield  {journal} {\bibinfo  {journal} {Nat. Phys.}\ }\textbf {\bibinfo {volume} {12}},\ \bibinfo {pages} {323} (\bibinfo {year} {2016})}\BibitemShut {NoStop}%
\bibitem [{\citenamefont {Rodriguez}(2016)}]{10.1088/0143-0807/37/2/025802}%
  \BibitemOpen
  \bibfield  {author} {\bibinfo {author} {\bibfnamefont {S.~R.-K.}\ \bibnamefont {Rodriguez}},\ }\bibfield  {title} {\bibinfo {title} {Classical and quantum distinctions between weak and strong coupling},\ }\href {https://doi.org/10.1088/0143-0807/37/2/025802} {\bibfield  {journal} {\bibinfo  {journal} {Eur. J. Phys.}\ }\textbf {\bibinfo {volume} {37}},\ \bibinfo {pages} {025802} (\bibinfo {year} {2016})}\BibitemShut {NoStop}%
\bibitem [{\citenamefont {Fitzpatrick}(2015)}]{fitzpatrick2015}%
  \BibitemOpen
  \bibfield  {author} {\bibinfo {author} {\bibfnamefont {R.}~\bibnamefont {Fitzpatrick}},\ }\href {https://books.google.com/books?id=v947DQAAQBAJ} {\emph {\bibinfo {title} {Quantum Mechanics}}}\ (\bibinfo  {publisher} {World Scientific Publishing Company},\ \bibinfo {year} {2015})\BibitemShut {NoStop}%
\bibitem [{\citenamefont {Mitterreiter}\ \emph {et~al.}(2021)\citenamefont {Mitterreiter}, \citenamefont {Schuler}, \citenamefont {Micevic}, \citenamefont {Hernang{\'o}mez-P{\'e}rez}, \citenamefont {Barthelmi}, \citenamefont {Cochrane}, \citenamefont {Kiemle}, \citenamefont {Sigger}, \citenamefont {Klein}, \citenamefont {Wong}, \citenamefont {Barnard}, \citenamefont {Watanabe}, \citenamefont {Taniguchi}, \citenamefont {Lorke}, \citenamefont {Jahnke}, \citenamefont {Finley}, \citenamefont {Schwartzberg}, \citenamefont {Qiu}, \citenamefont {Refaely-Abramson}, \citenamefont {Holleitner}, \citenamefont {Weber-Bargioni},\ and\ \citenamefont {Kastl}}]{10.1038/s41467-021-24102-y}%
  \BibitemOpen
  \bibfield  {author} {\bibinfo {author} {\bibfnamefont {E.}~\bibnamefont {Mitterreiter}}, \bibinfo {author} {\bibfnamefont {B.}~\bibnamefont {Schuler}}, \bibinfo {author} {\bibfnamefont {A.}~\bibnamefont {Micevic}}, \bibinfo {author} {\bibfnamefont {D.}~\bibnamefont {Hernang{\'o}mez-P{\'e}rez}}, \bibinfo {author} {\bibfnamefont {K.}~\bibnamefont {Barthelmi}}, \bibinfo {author} {\bibfnamefont {K.~A.}\ \bibnamefont {Cochrane}}, \bibinfo {author} {\bibfnamefont {J.}~\bibnamefont {Kiemle}}, \bibinfo {author} {\bibfnamefont {F.}~\bibnamefont {Sigger}}, \bibinfo {author} {\bibfnamefont {J.}~\bibnamefont {Klein}}, \bibinfo {author} {\bibfnamefont {E.}~\bibnamefont {Wong}}, \bibinfo {author} {\bibfnamefont {E.~S.}\ \bibnamefont {Barnard}}, \bibinfo {author} {\bibfnamefont {K.}~\bibnamefont {Watanabe}}, \bibinfo {author} {\bibfnamefont {T.}~\bibnamefont {Taniguchi}}, \bibinfo {author} {\bibfnamefont {M.}~\bibnamefont {Lorke}}, \bibinfo {author} {\bibfnamefont {F.}~\bibnamefont {Jahnke}}, \bibinfo {author} {\bibfnamefont {J.~J.}\ \bibnamefont {Finley}}, \bibinfo {author} {\bibfnamefont {A.~M.}\ \bibnamefont {Schwartzberg}}, \bibinfo {author} {\bibfnamefont {D.~Y.}\ \bibnamefont {Qiu}}, \bibinfo {author} {\bibfnamefont {S.}~\bibnamefont {Refaely-Abramson}}, \bibinfo {author} {\bibfnamefont {A.~W.}\ \bibnamefont {Holleitner}}, \bibinfo {author} {\bibfnamefont {A.}~\bibnamefont {Weber-Bargioni}},\ and\ \bibinfo {author} {\bibfnamefont {C.}~\bibnamefont {Kastl}},\ }\bibfield  {title} {\bibinfo {title} {The role of chalcogen vacancies for atomic defect emission in {MoS2}},\ }\href {https://doi.org/10.1038/s41467-021-24102-y} {\bibfield  {journal} {\bibinfo  {journal} {Nat. Commun.}\ }\textbf {\bibinfo {volume} {12}},\ \bibinfo {pages} {3822} (\bibinfo {year} {2021})}\BibitemShut {NoStop}%
\bibitem [{\citenamefont {Kaasbjerg}\ \emph {et~al.}(2012)\citenamefont {Kaasbjerg}, \citenamefont {Thygesen},\ and\ \citenamefont {Jacobsen}}]{PhysRevB.85.115317}%
  \BibitemOpen
  \bibfield  {author} {\bibinfo {author} {\bibfnamefont {K.}~\bibnamefont {Kaasbjerg}}, \bibinfo {author} {\bibfnamefont {K.~S.}\ \bibnamefont {Thygesen}},\ and\ \bibinfo {author} {\bibfnamefont {K.~W.}\ \bibnamefont {Jacobsen}},\ }\bibfield  {title} {\bibinfo {title} {Phonon-limited mobility in $n$-type single-layer {MoS${}_{2}$} from first principles},\ }\href {https://doi.org/10.1103/PhysRevB.85.115317} {\bibfield  {journal} {\bibinfo  {journal} {Phys. Rev. B}\ }\textbf {\bibinfo {volume} {85}},\ \bibinfo {pages} {115317} (\bibinfo {year} {2012})}\BibitemShut {NoStop}%
\bibitem [{\citenamefont {Qiu}\ \emph {et~al.}(2015)\citenamefont {Qiu}, \citenamefont {Cao},\ and\ \citenamefont {Louie}}]{PhysRevLett.115.176801}%
  \BibitemOpen
  \bibfield  {author} {\bibinfo {author} {\bibfnamefont {D.~Y.}\ \bibnamefont {Qiu}}, \bibinfo {author} {\bibfnamefont {T.}~\bibnamefont {Cao}},\ and\ \bibinfo {author} {\bibfnamefont {S.~G.}\ \bibnamefont {Louie}},\ }\bibfield  {title} {\bibinfo {title} {Nonanalyticity, valley quantum phases, and lightlike exciton dispersion in monolayer transition metal dichalcogenides: Theory and first-principles calculations},\ }\href {https://doi.org/10.1103/PhysRevLett.115.176801} {\bibfield  {journal} {\bibinfo  {journal} {Phys. Rev. Lett.}\ }\textbf {\bibinfo {volume} {115}},\ \bibinfo {pages} {176801} (\bibinfo {year} {2015})}\BibitemShut {NoStop}%
\bibitem [{\citenamefont {Chow}\ \emph {et~al.}(2017)\citenamefont {Chow}, \citenamefont {Yu}, \citenamefont {Jones}, \citenamefont {Schaibley}, \citenamefont {Koehler}, \citenamefont {Mandrus}, \citenamefont {Merlin}, \citenamefont {Yao},\ and\ \citenamefont {Xu}}]{10.1038/s41699-017-0035-1}%
  \BibitemOpen
  \bibfield  {author} {\bibinfo {author} {\bibfnamefont {C.~M.}\ \bibnamefont {Chow}}, \bibinfo {author} {\bibfnamefont {H.}~\bibnamefont {Yu}}, \bibinfo {author} {\bibfnamefont {A.~M.}\ \bibnamefont {Jones}}, \bibinfo {author} {\bibfnamefont {J.~R.}\ \bibnamefont {Schaibley}}, \bibinfo {author} {\bibfnamefont {M.}~\bibnamefont {Koehler}}, \bibinfo {author} {\bibfnamefont {D.~G.}\ \bibnamefont {Mandrus}}, \bibinfo {author} {\bibfnamefont {R.}~\bibnamefont {Merlin}}, \bibinfo {author} {\bibfnamefont {W.}~\bibnamefont {Yao}},\ and\ \bibinfo {author} {\bibfnamefont {X.}~\bibnamefont {Xu}},\ }\bibfield  {title} {\bibinfo {title} {Phonon-assisted oscillatory exciton dynamics in monolayer {MoSe2}},\ }\href {https://doi.org/10.1038/s41699-017-0035-1} {\bibfield  {journal} {\bibinfo  {journal} {npj 2D Mater. and Appl.}\ }\textbf {\bibinfo {volume} {1}},\ \bibinfo {pages} {33} (\bibinfo {year} {2017})}\BibitemShut {NoStop}%
\bibitem [{\citenamefont {Bergh\"auser}\ \emph {et~al.}(2018)\citenamefont {Bergh\"auser}, \citenamefont {Steinleitner}, \citenamefont {Merkl}, \citenamefont {Huber}, \citenamefont {Knorr},\ and\ \citenamefont {Malic}}]{PhysRevB.98.020301}%
  \BibitemOpen
  \bibfield  {author} {\bibinfo {author} {\bibfnamefont {G.}~\bibnamefont {Bergh\"auser}}, \bibinfo {author} {\bibfnamefont {P.}~\bibnamefont {Steinleitner}}, \bibinfo {author} {\bibfnamefont {P.}~\bibnamefont {Merkl}}, \bibinfo {author} {\bibfnamefont {R.}~\bibnamefont {Huber}}, \bibinfo {author} {\bibfnamefont {A.}~\bibnamefont {Knorr}},\ and\ \bibinfo {author} {\bibfnamefont {E.}~\bibnamefont {Malic}},\ }\bibfield  {title} {\bibinfo {title} {Mapping of the dark exciton landscape in transition metal dichalcogenides},\ }\href {https://doi.org/10.1103/PhysRevB.98.020301} {\bibfield  {journal} {\bibinfo  {journal} {Phys. Rev. B}\ }\textbf {\bibinfo {volume} {98}},\ \bibinfo {pages} {020301} (\bibinfo {year} {2018})}\BibitemShut {NoStop}%
\bibitem [{\citenamefont {Chi}\ \emph {et~al.}(2023)\citenamefont {Chi}, \citenamefont {Wei}, \citenamefont {Zhang}, \citenamefont {Chen},\ and\ \citenamefont {Weng}}]{10.1021/acs.jpclett.3c02431}%
  \BibitemOpen
  \bibfield  {author} {\bibinfo {author} {\bibfnamefont {Z.}~\bibnamefont {Chi}}, \bibinfo {author} {\bibfnamefont {Z.}~\bibnamefont {Wei}}, \bibinfo {author} {\bibfnamefont {G.}~\bibnamefont {Zhang}}, \bibinfo {author} {\bibfnamefont {H.}~\bibnamefont {Chen}},\ and\ \bibinfo {author} {\bibfnamefont {Y.-X.}\ \bibnamefont {Weng}},\ }\bibfield  {title} {\bibinfo {title} {Determining band splitting and spin-flip dynamics in monolayer {MoS2}},\ }\href {https://doi.org/10.1021/acs.jpclett.3c02431} {\bibfield  {journal} {\bibinfo  {journal} {J. Phys. Chem. Lett.}\ }\textbf {\bibinfo {volume} {14}},\ \bibinfo {pages} {9640} (\bibinfo {year} {2023})}\BibitemShut {NoStop}%
\bibitem [{\citenamefont {Zhang}\ \emph {et~al.}(2015)\citenamefont {Zhang}, \citenamefont {You}, \citenamefont {Zhao},\ and\ \citenamefont {Heinz}}]{PhysRevLett.115.257403}%
  \BibitemOpen
  \bibfield  {author} {\bibinfo {author} {\bibfnamefont {X.-X.}\ \bibnamefont {Zhang}}, \bibinfo {author} {\bibfnamefont {Y.}~\bibnamefont {You}}, \bibinfo {author} {\bibfnamefont {S.~Y.~F.}\ \bibnamefont {Zhao}},\ and\ \bibinfo {author} {\bibfnamefont {T.~F.}\ \bibnamefont {Heinz}},\ }\bibfield  {title} {\bibinfo {title} {Experimental evidence for dark excitons in monolayer {${\mathrm{WSe}}_{2}$}},\ }\href {https://doi.org/10.1103/PhysRevLett.115.257403} {\bibfield  {journal} {\bibinfo  {journal} {Phys. Rev. Lett.}\ }\textbf {\bibinfo {volume} {115}},\ \bibinfo {pages} {257403} (\bibinfo {year} {2015})}\BibitemShut {NoStop}%
\bibitem [{\citenamefont {Robert}\ \emph {et~al.}(2020)\citenamefont {Robert}, \citenamefont {Han}, \citenamefont {Kapuscinski}, \citenamefont {Delhomme}, \citenamefont {Faugeras}, \citenamefont {Amand}, \citenamefont {Molas}, \citenamefont {Bartos}, \citenamefont {Watanabe}, \citenamefont {Taniguchi}, \citenamefont {Urbaszek}, \citenamefont {Potemski},\ and\ \citenamefont {Marie}}]{10.1038/s41467-020-17608-4}%
  \BibitemOpen
  \bibfield  {author} {\bibinfo {author} {\bibfnamefont {C.}~\bibnamefont {Robert}}, \bibinfo {author} {\bibfnamefont {B.}~\bibnamefont {Han}}, \bibinfo {author} {\bibfnamefont {P.}~\bibnamefont {Kapuscinski}}, \bibinfo {author} {\bibfnamefont {A.}~\bibnamefont {Delhomme}}, \bibinfo {author} {\bibfnamefont {C.}~\bibnamefont {Faugeras}}, \bibinfo {author} {\bibfnamefont {T.}~\bibnamefont {Amand}}, \bibinfo {author} {\bibfnamefont {M.~R.}\ \bibnamefont {Molas}}, \bibinfo {author} {\bibfnamefont {M.}~\bibnamefont {Bartos}}, \bibinfo {author} {\bibfnamefont {K.}~\bibnamefont {Watanabe}}, \bibinfo {author} {\bibfnamefont {T.}~\bibnamefont {Taniguchi}}, \bibinfo {author} {\bibfnamefont {B.}~\bibnamefont {Urbaszek}}, \bibinfo {author} {\bibfnamefont {M.}~\bibnamefont {Potemski}},\ and\ \bibinfo {author} {\bibfnamefont {X.}~\bibnamefont {Marie}},\ }\bibfield  {title} {\bibinfo {title} {{Measurement of the spin-forbidden dark excitons in MoS2 and MoSe2 monolayers}},\ }\href {https://doi.org/10.1038/s41467-020-17608-4} {\bibfield  {journal} {\bibinfo  {journal} {Nat. Commun.}\ }\textbf {\bibinfo {volume} {11}},\ \bibinfo {pages} {4037} (\bibinfo {year} {2020})}\BibitemShut {NoStop}%
\bibitem [{\citenamefont {Yu}\ \emph {et~al.}(2019)\citenamefont {Yu}, \citenamefont {Laurien}, \citenamefont {Hu},\ and\ \citenamefont {Rubel}}]{PhysRevB.100.125413}%
  \BibitemOpen
  \bibfield  {author} {\bibinfo {author} {\bibfnamefont {H.}~\bibnamefont {Yu}}, \bibinfo {author} {\bibfnamefont {M.}~\bibnamefont {Laurien}}, \bibinfo {author} {\bibfnamefont {Z.}~\bibnamefont {Hu}},\ and\ \bibinfo {author} {\bibfnamefont {O.}~\bibnamefont {Rubel}},\ }\bibfield  {title} {\bibinfo {title} {{Exploration of the bright and dark exciton landscape and fine structure of ${\mathrm{MoS}}_{2}$ using ${\mathrm{G}}_{0}{\mathrm{W}}_{0}$-BSE}},\ }\href {https://doi.org/10.1103/PhysRevB.100.125413} {\bibfield  {journal} {\bibinfo  {journal} {Phys. Rev. B}\ }\textbf {\bibinfo {volume} {100}},\ \bibinfo {pages} {125413} (\bibinfo {year} {2019})}\BibitemShut {NoStop}%
\bibitem [{\citenamefont {Feierabend}\ \emph {et~al.}(2020)\citenamefont {Feierabend}, \citenamefont {Brem}, \citenamefont {Ekman},\ and\ \citenamefont {Malic}}]{10.1088/2053-1583/abb876}%
  \BibitemOpen
  \bibfield  {author} {\bibinfo {author} {\bibfnamefont {M.}~\bibnamefont {Feierabend}}, \bibinfo {author} {\bibfnamefont {S.}~\bibnamefont {Brem}}, \bibinfo {author} {\bibfnamefont {A.}~\bibnamefont {Ekman}},\ and\ \bibinfo {author} {\bibfnamefont {E.}~\bibnamefont {Malic}},\ }\bibfield  {title} {\bibinfo {title} {Brightening of spin- and momentum-dark excitons in transition metal dichalcogenides},\ }\href {https://doi.org/10.1088/2053-1583/abb876} {\bibfield  {journal} {\bibinfo  {journal} {2D Mater.}\ }\textbf {\bibinfo {volume} {8}},\ \bibinfo {pages} {015013} (\bibinfo {year} {2020})}\BibitemShut {NoStop}%
\bibitem [{\citenamefont {Feierabend}\ \emph {et~al.}(2017)\citenamefont {Feierabend}, \citenamefont {Bergh{\"a}user}, \citenamefont {Knorr},\ and\ \citenamefont {Malic}}]{10.1038/ncomms14776}%
  \BibitemOpen
  \bibfield  {author} {\bibinfo {author} {\bibfnamefont {M.}~\bibnamefont {Feierabend}}, \bibinfo {author} {\bibfnamefont {G.}~\bibnamefont {Bergh{\"a}user}}, \bibinfo {author} {\bibfnamefont {A.}~\bibnamefont {Knorr}},\ and\ \bibinfo {author} {\bibfnamefont {E.}~\bibnamefont {Malic}},\ }\bibfield  {title} {\bibinfo {title} {Proposal for dark exciton based chemical sensors},\ }\href {https://doi.org/10.1038/ncomms14776} {\bibfield  {journal} {\bibinfo  {journal} {Nat. Commun.}\ }\textbf {\bibinfo {volume} {8}},\ \bibinfo {pages} {14776} (\bibinfo {year} {2017})}\BibitemShut {NoStop}%
\bibitem [{\citenamefont {Khatibi}\ \emph {et~al.}(2018)\citenamefont {Khatibi}, \citenamefont {Feierabend}, \citenamefont {Selig}, \citenamefont {Brem}, \citenamefont {Linderälv}, \citenamefont {Erhart},\ and\ \citenamefont {Malic}}]{10.1088/2053-1583/aae953}%
  \BibitemOpen
  \bibfield  {author} {\bibinfo {author} {\bibfnamefont {Z.}~\bibnamefont {Khatibi}}, \bibinfo {author} {\bibfnamefont {M.}~\bibnamefont {Feierabend}}, \bibinfo {author} {\bibfnamefont {M.}~\bibnamefont {Selig}}, \bibinfo {author} {\bibfnamefont {S.}~\bibnamefont {Brem}}, \bibinfo {author} {\bibfnamefont {C.}~\bibnamefont {Linderälv}}, \bibinfo {author} {\bibfnamefont {P.}~\bibnamefont {Erhart}},\ and\ \bibinfo {author} {\bibfnamefont {E.}~\bibnamefont {Malic}},\ }\bibfield  {title} {\bibinfo {title} {Impact of strain on the excitonic linewidth in transition metal dichalcogenides},\ }\href {https://doi.org/10.1088/2053-1583/aae953} {\bibfield  {journal} {\bibinfo  {journal} {2D Mater.}\ }\textbf {\bibinfo {volume} {6}},\ \bibinfo {pages} {015015} (\bibinfo {year} {2018})}\BibitemShut {NoStop}%
\bibitem [{\citenamefont {Malic}\ \emph {et~al.}(2018)\citenamefont {Malic}, \citenamefont {Selig}, \citenamefont {Feierabend}, \citenamefont {Brem}, \citenamefont {Christiansen}, \citenamefont {Wendler}, \citenamefont {Knorr},\ and\ \citenamefont {Bergh\"auser}}]{PhysRevMaterials.2.014002}%
  \BibitemOpen
  \bibfield  {author} {\bibinfo {author} {\bibfnamefont {E.}~\bibnamefont {Malic}}, \bibinfo {author} {\bibfnamefont {M.}~\bibnamefont {Selig}}, \bibinfo {author} {\bibfnamefont {M.}~\bibnamefont {Feierabend}}, \bibinfo {author} {\bibfnamefont {S.}~\bibnamefont {Brem}}, \bibinfo {author} {\bibfnamefont {D.}~\bibnamefont {Christiansen}}, \bibinfo {author} {\bibfnamefont {F.}~\bibnamefont {Wendler}}, \bibinfo {author} {\bibfnamefont {A.}~\bibnamefont {Knorr}},\ and\ \bibinfo {author} {\bibfnamefont {G.}~\bibnamefont {Bergh\"auser}},\ }\bibfield  {title} {\bibinfo {title} {Dark excitons in transition metal dichalcogenides},\ }\href {https://doi.org/10.1103/PhysRevMaterials.2.014002} {\bibfield  {journal} {\bibinfo  {journal} {Phys. Rev. Mater.}\ }\textbf {\bibinfo {volume} {2}},\ \bibinfo {pages} {014002} (\bibinfo {year} {2018})}\BibitemShut {NoStop}%
\bibitem [{\citenamefont {Qian}\ \emph {et~al.}(2022{\natexlab{a}})\citenamefont {Qian}, \citenamefont {Villafa\~ne}, \citenamefont {Soubelet}, \citenamefont {H\"otger}, \citenamefont {Taniguchi}, \citenamefont {Watanabe}, \citenamefont {Wilson}, \citenamefont {Stier}, \citenamefont {Holleitner},\ and\ \citenamefont {Finley}}]{PhysRevLett.128.237403}%
  \BibitemOpen
  \bibfield  {author} {\bibinfo {author} {\bibfnamefont {C.}~\bibnamefont {Qian}}, \bibinfo {author} {\bibfnamefont {V.}~\bibnamefont {Villafa\~ne}}, \bibinfo {author} {\bibfnamefont {P.}~\bibnamefont {Soubelet}}, \bibinfo {author} {\bibfnamefont {A.}~\bibnamefont {H\"otger}}, \bibinfo {author} {\bibfnamefont {T.}~\bibnamefont {Taniguchi}}, \bibinfo {author} {\bibfnamefont {K.}~\bibnamefont {Watanabe}}, \bibinfo {author} {\bibfnamefont {N.~P.}\ \bibnamefont {Wilson}}, \bibinfo {author} {\bibfnamefont {A.~V.}\ \bibnamefont {Stier}}, \bibinfo {author} {\bibfnamefont {A.~W.}\ \bibnamefont {Holleitner}},\ and\ \bibinfo {author} {\bibfnamefont {J.~J.}\ \bibnamefont {Finley}},\ }\bibfield  {title} {\bibinfo {title} {Nonlocal exciton-photon interactions in hybrid high-{$Q$} beam nanocavities with encapsulated {${\mathrm{MoS}}_{2}$} monolayers},\ }\href {https://doi.org/10.1103/PhysRevLett.128.237403} {\bibfield  {journal} {\bibinfo  {journal} {Phys. Rev. Lett.}\ }\textbf {\bibinfo {volume} {128}},\ \bibinfo {pages} {237403} (\bibinfo {year} {2022}{\natexlab{a}})}\BibitemShut {NoStop}%
\bibitem [{\citenamefont {Eichenfield}\ \emph {et~al.}(2009)\citenamefont {Eichenfield}, \citenamefont {Chan}, \citenamefont {Camacho}, \citenamefont {Vahala},\ and\ \citenamefont {Painter}}]{10.1038/nature08524}%
  \BibitemOpen
  \bibfield  {author} {\bibinfo {author} {\bibfnamefont {M.}~\bibnamefont {Eichenfield}}, \bibinfo {author} {\bibfnamefont {J.}~\bibnamefont {Chan}}, \bibinfo {author} {\bibfnamefont {R.~M.}\ \bibnamefont {Camacho}}, \bibinfo {author} {\bibfnamefont {K.~J.}\ \bibnamefont {Vahala}},\ and\ \bibinfo {author} {\bibfnamefont {O.}~\bibnamefont {Painter}},\ }\bibfield  {title} {\bibinfo {title} {Optomechanical crystals},\ }\href {https://doi.org/10.1038/nature08524} {\bibfield  {journal} {\bibinfo  {journal} {Nature}\ }\textbf {\bibinfo {volume} {462}},\ \bibinfo {pages} {78} (\bibinfo {year} {2009})}\BibitemShut {NoStop}%
\bibitem [{\citenamefont {Stiehm}\ \emph {et~al.}(2019)\citenamefont {Stiehm}, \citenamefont {Schneider}, \citenamefont {Kern}, \citenamefont {Niehues}, \citenamefont {Michaelis~de Vasconcellos},\ and\ \citenamefont {Bratschitsch}}]{10.1063/1.5100593}%
  \BibitemOpen
  \bibfield  {author} {\bibinfo {author} {\bibfnamefont {T.}~\bibnamefont {Stiehm}}, \bibinfo {author} {\bibfnamefont {R.}~\bibnamefont {Schneider}}, \bibinfo {author} {\bibfnamefont {J.}~\bibnamefont {Kern}}, \bibinfo {author} {\bibfnamefont {I.}~\bibnamefont {Niehues}}, \bibinfo {author} {\bibfnamefont {S.}~\bibnamefont {Michaelis~de Vasconcellos}},\ and\ \bibinfo {author} {\bibfnamefont {R.}~\bibnamefont {Bratschitsch}},\ }\bibfield  {title} {\bibinfo {title} {Supercontinuum second harmonic generation spectroscopy of atomically thin semiconductors},\ }\href {https://doi.org/10.1063/1.5100593} {\bibfield  {journal} {\bibinfo  {journal} {Rev. Sci. Instrum.}\ }\textbf {\bibinfo {volume} {90}},\ \bibinfo {pages} {083102} (\bibinfo {year} {2019})}\BibitemShut {NoStop}%
\bibitem [{\citenamefont {Qian}\ \emph {et~al.}(2022{\natexlab{b}})\citenamefont {Qian}, \citenamefont {Villafañe}, \citenamefont {Schalk}, \citenamefont {Astakhov}, \citenamefont {Kentsch}, \citenamefont {Helm}, \citenamefont {Soubelet}, \citenamefont {Stier},\ and\ \citenamefont {Finley}}]{2210.00150}%
  \BibitemOpen
  \bibfield  {author} {\bibinfo {author} {\bibfnamefont {C.}~\bibnamefont {Qian}}, \bibinfo {author} {\bibfnamefont {V.}~\bibnamefont {Villafañe}}, \bibinfo {author} {\bibfnamefont {M.}~\bibnamefont {Schalk}}, \bibinfo {author} {\bibfnamefont {G.~V.}\ \bibnamefont {Astakhov}}, \bibinfo {author} {\bibfnamefont {U.}~\bibnamefont {Kentsch}}, \bibinfo {author} {\bibfnamefont {M.}~\bibnamefont {Helm}}, \bibinfo {author} {\bibfnamefont {P.}~\bibnamefont {Soubelet}}, \bibinfo {author} {\bibfnamefont {A.~V.}\ \bibnamefont {Stier}},\ and\ \bibinfo {author} {\bibfnamefont {J.~J.}\ \bibnamefont {Finley}},\ }\href@noop {} {\bibinfo {title} {Emitter-optomechanical interaction in ultra-high-{Q} hbn nanocavities}} (\bibinfo {year} {2022}{\natexlab{b}}),\ \Eprint {https://arxiv.org/abs/arXiv:2210.00150} {arXiv:2210.00150} \BibitemShut {NoStop}%
\bibitem [{\citenamefont {Hoch}\ \emph {et~al.}(2022)\citenamefont {Hoch}, \citenamefont {Yao},\ and\ \citenamefont {Poot}}]{10.1021/acs.nanolett.2c00613}%
  \BibitemOpen
  \bibfield  {author} {\bibinfo {author} {\bibfnamefont {D.}~\bibnamefont {Hoch}}, \bibinfo {author} {\bibfnamefont {X.}~\bibnamefont {Yao}},\ and\ \bibinfo {author} {\bibfnamefont {M.}~\bibnamefont {Poot}},\ }\bibfield  {title} {\bibinfo {title} {Geometric tuning of stress in predisplaced silicon nitride resonators},\ }\href {https://doi.org/10.1021/acs.nanolett.2c00613} {\bibfield  {journal} {\bibinfo  {journal} {Nano Lett.}\ }\textbf {\bibinfo {volume} {22}},\ \bibinfo {pages} {4013} (\bibinfo {year} {2022})}\BibitemShut {NoStop}%
\end{thebibliography}
